\documentclass[a4paper,twocolumn,10pt]{IEEEtran}

\usepackage[usenames, dvipsnames]{color}
\usepackage[cmex10]{amsmath} 
\usepackage{amssymb,amsthm}

\usepackage{tikzexternal}
\usepackage{calc}
\usepackage{cite}

\newcommand{\fm}[1]{\scriptsize\mbox{\ensuremath{#1}}}
\newcommand{\ft}[1]{\scriptsize #1}

\newtheorem{theorem}{Theorem}
\newtheorem{lemma}{Lemma}
\newtheorem{corollary}{Corollary}

\newcommand{\ie}{{\it i.e.},\ }
\newcommand{\eg}{{\it e.g.},\ }
\newcommand{\cf}{{\it cf.} }

\newcommand{\FF}{\mathbb{F}}

\newcommand{\LL}{\mathcal{L}}

\newcommand{\Rnet}{R_{\mathrm{net}}}
\newcommand{\barRnet}{\overline{R}_{\mathrm{net}}}
\newcommand{\Rresolve}{R_{\mathrm{res}}}
\newcommand{\barRresolve}{\overline{R}_{\mathrm{res}}}
\newcommand{\barRresolveSIC}{\overline{R}_{\mathrm{res, SIC}}}

\newcommand{\Rplnc}{R_{\mathrm{plnc}}}

\newcommand{\JG}[1]{#1}
\newcommand{\JGb}[1]{#1}
\newcommand{\CS}[1]{#1}
\newcommand{\CSm}[1]{#1}

\newcommand{\JGc}[1]{#1}
\newcommand{\JGf}[1]{#1}
\newcommand{\PP}[1]{#1}
\newcommand{\CSrev}[1]{#1}

\begin{document}

\title{Sign-Compute-Resolve for Tree Splitting\\ Random Access}

\author{
Jasper Goseling, \v Cedomir Stefanovi\' c and Petar Popovski
\thanks{
The work of J. Goseling was supported in part by the Netherlands Organization for Scientific Research (NWO), grant $612.001.107$.
The work of \v C. Stefanovi\' c was supported by the Danish Council for Independent Research (DFF),  grant DFF-4005-00281.
The work of P. Popovski has been in part supported by the European
Research Council (ERC Consolidator Grant nr. 648382 WILLOW)
within the Horizon 2020 Program.

Part of this work was presented at the IEEE International Conference on Communications, 2014 and at the 52nd Annual Allerton Conference on Communication, Control, and Computing, 2014.

J.\ Goseling is with the Department of Applied Mathematics, University of Twente, Drienerlolaan 5, 7522 NB Enschede, The Netherlands. (email: j.goseling@utwente.nl)

\v C.\  Stefanovi\' c and P.\ Popovski are with the Department of Electronic Systems, Aalborg University, Denmark. (email: cs@es.aau.dk, petarp@es.aau.dk)

}
}


\maketitle

%
%
%
\begin{abstract}
We present a \PP{framework for} random access that is based on three elements: physical-layer network coding (PLNC), signature codes and tree splitting.
In presence of a collision, physical-layer network coding enables the receiver to decode, \CS{\ie compute}, the sum of the \CS{packets} that were transmitted by the individual users.
For each user, the packet consists of \JGb{the} user's signature, as well as the data that the user wants to communicate.
As long as no more than $K$ users collide, their identities can be recovered from the sum of their signatures.
\PP{This framework for creating and transmitting packets can be used as a fundamental building block in random access algorithms, since it helps to deal efficiently with the uncertainty of the set of contending terminals. In this paper we show how to apply the 
framework in conjunction with a tree-splitting algorithm, which is required to} deal with the case \JGb{that} more than $K$ users collide. 
We demonstrate that our approach achieves throughput that tends to 1 rapidly as $K$ increases.
We also present results on net data-rate of the system, showing the impact of the overheads of the constituent elements of the proposed protocol.
\JGb{We compare the performance of our scheme with an upper bound that is obtained under the assumption that the active users are a priori known. Also, we consider an upper bound on the net data-rate for any PLNC based strategy in which one linear equation per slot is decoded.}
\CSm{We show that already at modest packet lengths, the net data-rate of our scheme becomes close to the second upper bound, \ie} the overhead of the \JGb{contention resolution algorithm and the signature codes vanishes.}
\end{abstract}

%
%
\section{Introduction} \label{sec:intro}

%
%


Uncertainty is the essential element of communication systems, caused by noise, errors, and random traffic (packet) arrivals at user(s).
A canonical example of the latter is seen in random access protocols, used for handling transmissions of users to a common receiver, \eg a base station, over a shared wireless medium. 
Random access is necessary when the total number of users associated with the base station is large, but \JGb{within} a given short time interval, the number of active users that have packets to transmit is small and unknown {\em a priori}.
Such is the case \JG{in, for instance, wide-area} networks of sensors, where each sensor has a sporadic traffic pattern.
The goal of random access protocols is to enable each of the active users to eventually send its packet successfully. 

Traditionally, random access protocols have been designed under the \emph{collision model}: when two or more users transmit at the same time, a collision occurs and all involved transmissions are lost.
In other words, collisions are considered as destructive and the information contained in them as irrecoverable.
Therefore, the objective of classical random access protocols, such as ALOHA \cite{R1975} or tree splitting \cite{capetanakis}, is to ensure that each user gets the opportunity to send its packet without collision.
On the other hand, the recent inclusion of elaborate physical layer techniques in random access protocols allows for the extension of the design space beyond the collision model\cite{NDMA,SICTAb,goseling2013physical,goseling2015access_it,PSLP2014}, such that collisions are treated as sums of packets and, instead of being discarded, they are buffered and reused.
We motivate the benefits of such an approach through a simple example, where the received signals in the first two slots are
\begin{align}
\label{eq:SimpleSIC}
Y_1 & = X_1 + X_2 + Z_1, \nonumber \\
Y_2 & = X_2 + Z_2,
\end{align}
where $X_1$ and $X_2$ are the user signals (packets) and $Z_1$ and $Z_2$ represent the noise. 
The received signal $Y_1$ is buffered, and, if $X_2$ is successfully decoded from slot $Y_2$, it can be subtracted (\ie cancelled) from $Y_1$.
The receiver proceeds by attempting to decode $X_1$ from the noisy signal $ Y_1 - X_2 = X_1 + Z$.
Obviously, the exploitation of the information contained in the collision slot $Y_1$ boosts the protocol performance.


The motivating example also demonstrates that, in the general framework, the impact of the noise cannot be neglected.
This is fundamentally changed by applying reliable physical layer network coding (PLNC) to the problem of random access.
The key idea in PLNC is to decode a function of multiple received signals, rather than decoding the individual signals.
Such operation is termed denoise-and-forward \cite{popovski2007physical,popovski2006anti}, or compute-and-forward \cite{nazer11compforw}, \CS{the latter being the motivation \JGb{for} part of the name of the scheme proposed in the paper.}
Assume that $W_1$ and $W_2$ represent the data as a sequence of symbols from finite field $\FF_q$ that are mapped to the baseband signals $X_1$ and $X_2$\JGb{, respectively, in} example (\ref{eq:SimpleSIC}).
Upon receiving $Y_1$ from (\ref{eq:SimpleSIC}), the base station stores $W_1 + W_2$. If $X_2$ (\ie $W_2$) is decoded from $Y_2$, then $W_1$ can be obtained from the stored signal $W_1 + W_2$, \ie the sum in $\FF_q$ of $W_1$ and $W_2$.
In brief, the use of PLNC removes the uncertainty of the noise, \CS{leaving the receiver only} with the uncertainty about the contending set of users. 


\JG{
One of the main challenges in the application of PLNC in random access protocols is for the receiver to learn the set of transmitting users~\cite{goseling2015access_it}. In~\eqref{eq:SimpleSIC}, for instance, the receiver does not know that $X_1$ is sent in slots $1$ and $2$, and $X_2$ in slot $2$. This sets the motivation to introduce PLNC-based random access with signatures in~\cite{goseling14massap, GSP2014}, where the users \JGc{prepend to their messages} a codeword of a $K$-out-of-$N$ code~\cite{M1990, jevtic1995families, gyorfi2004signature}. More precisely, the $\ell-$th user applies the following communication strategy: it prepends a \emph{signature} $W^s_\ell$, consisting of predefined number of symbols, to the message $W_\ell^d$ in order to obtain $W_\ell$. The signature is based on a code that has the following property: if at most $K$ users transmit in a given slot, then from the sum of the signatures 
the receiver knows exactly which transmitters have contributed to the data stored in the present slot. In other words, the sum
\begin{align} \label{eq:SignatureSum}
\sum_{\ell = 1}^{L} W^s_{\ell}
\end{align}
is uniquely decodable if $L \leq K$ \CS{, where $L$ denotes the collision multiplicity}.}


\PP{The addition of signatures to packets and its combination with PLNC brings a conceptual novelty in the design of random access algorithms. The reason is that, when the number of colliding terminals is at most $K$, then the receiver can find out the identities of those terminals, but it cannot decode the data of all of them. This means that from a collision with $K$ or less contenders, the receiver can extract protocol-related information, but not the data. This is a generalization of the common case when $K=1$, where a single-user transmission is required to decode both protocol information and data. The efficiency of this approach is shown in the context of} a tree-splitting algorithm for contention resolution. 
\PP{We emphasize that} the mechanisms enabled by the signatures and PLNC are not limited to tree-splitting algorithms, but we have selected the tree-splitting framework as it is known to provide the highest possible throughput for the traditional collision model \cite{Massey}. In fact, these mechanisms can be applied in any framework of multiple access protocols in which the collisions are not wasted, but used in decoding, as in \eg coded random access \cite{PSLP2014}.
The use of PLNC transforms the multiple access channel into an $\FF_q$ adder channel\CS{, \ie it removes the uncertainty due to noise}.
The addition of signature coding facilitates the generalization of the concept of collision, which is a conceptual shift \JGb{away from} the standard tree splitting context. 
Specifically, it allows the receiver to obtain (i) the knowledge of the collision multiplicity $L$, \ie the number of the collided user signals in the slot, and (ii) the resolution of collided user identities, if \CS{$L \leq K$}, where $K$ is a design parameter.
These features lead to a revision of the objective of contention resolution as compared to the standard schemes: to drive the contending users in a state in which the collision multiplicity becomes resolvable and the receiver is able to get a sufficient number of \emph{equations} in the finite field in order to be able to decode the users' data.
We analyze the proposed scheme and provide results related to the expected duration of the contention resolution interval, the expected number of resolved users per slot (\ie throughput) and the net rate of information transmission, which takes into account the overhead related to PLNC and signatures.
We show that the use of signatures is significantly reducing the average time required to extract useful information from the collisions, therefore improving the overall throughput performance.
On the other hand, the assessment of the net rate provides insights in the basic tradeoffs and mechanisms that need to be considered for a contention resolution based on PLNC and signatures.
To the best of our knowledge, this type of analysis is typically omitted in \CS{the literature on tree-splitting algorithms}. The reason is that in random access literature a packet is the atomic unit of communication, such that the overhead related to the contention identifiers and channel coding are not modeled and throughput is the primary performance parameter.
\JG{We compare the net rate of our scheme with upper bounds on the net rate.}
We also extend the scheme to include successive interference cancellation (SIC), enabling the use of collisions with multiplicities larger than $K$, and we provide the accompanying analysis and performance assessment.
Note that SIC can be seamlessly incorporated into the proposed framework, as \JG{the PLNC reduces the interference cancellation to simple operations in $\FF_q$.}
\CS{Finally, we note that the proposed scheme is simple to implement at the transmitter side, requiring use of a linear code and a unique signature.
This is an important criterion for scenarios in which a massive number of sensors (or other small devices) transmit information to a common receiver.
Our analysis of the achieved net rate, including the upper bounds, demonstrate that} keeping the complexity low results only in a small penalty.

The paper is organized as follows.
In Section~\ref{sec:related} we discuss related work.
In Section~\ref{sec:model} we introduce the model.
In Section~\ref{sec:prelim} we present results on PLNC, signature codes and tree splitting that will be used in the remainder of the paper.
The proposed strategy is presented in Section~\ref{sec:idea} and its performance analyzed in Section~\ref{sec:analysis}.
An extension of the strategy that involves successive interference cancellation is presented in Section~\ref{sec:sic}. Finally, a discussion and concluding remarks are given in Section~\ref{sec:discussion}.

\section{Related Work} \label{sec:related}

The use of PLNC for random access was studied in~\cite{parandehgheibi2010collision,parandehgheibi2010acknowledgement,cocco2011vtc, cocco2012arxiv,goseling2013random,goseling2013physical}, where it was assumed that the receiver knows which users are active in each slot.
\JG{
Other ways to relax this assumption that do not involve signatures are to rely on the successful decoding of signals from singleton slots~\cite{SICTAb,PSLP2014}, as suggested in example \eqref{eq:SimpleSIC}, or to rely on complex signal separation techniques \cite{NDMA}.}

The use of PLNC and signature codes was considered in~\cite{censor2012bounded} for broadcast in networks.
The combination of PLNC and signature codes for random access was introduced in~\cite{goseling14massap}.
 Both in~\cite{censor2012bounded} and~\cite{goseling14massap} it is assumed that the number of contending users is bounded.
 In this work we set aside this assumption and design a contention resolution algorithm that deals with any number of contending users.

A comprehensive review of signature coding and its application to multiple access can be found in \cite{MACbookb}.
The reviewed results mainly involve existence proofs of certain types of signature codes, leaving the contention resolution and, in general, random access protocol operation out of focus.
Two exceptions can be found in \cite{RM1993} and \cite{RM1994}, where the authors consider a tree-splitting and an ALOHA based random access solution, respectively, which exploit $K$-out-of-$M$ multiple access codes proposed in \cite{M1990}.
The approach suggested in \cite{RM1993} resembles the one proposed in the paper, however, the authors neglect the impact of noise, conclude that \JGb{a} choice of $K=3$ is optimal for the code construction from \cite{M1990}, and also show that the conventional tree splitting actually outperforms their solution in terms of user resolution rate in the case of blocked access, due to \JGb{the fact that} the code length is equal to $M$.
On the contrary, our approach is based on much shorter codes, whose length scales roughly as $K \log M$, the impact of noise is explicitly taken into account, and the performance is thoroughly characterized in terms of $K$, including insights in the choice of optimal values.

The scheme employed in the paper provides the receiver with the knowledge of the collision multiplicity.
Pippenger showed in a non-constructive way that this knowledge could be used to achieve throughputs that asymptotically tend to one under the collision channel model \cite{P1981}.
The construction of the protocol which leverages on these ideas was done in \cite{Ruszinko1997}, however, the proposed solution involves exponential computational complexity. 
On the other hand, the scheme employed in this paper achieves comparable performance, albeit using much simpler operating principles.

The analysis \JGb{of the tree-splitting algorithm} presented in \JGb{this} paper is based on the approach pioneered by Massey \cite{Massey}. 
The use of SIC in the contention resolution framework was first investigated in \cite{SICTAb}, where it was shown that throughput performance can be pushed to 0.693.
Another approach was suggested in \cite{SSP2013}, where SIC is employed over a set of partially split trees, and optimization was performed over the splitting strategy that favors fast SIC evolution.
The reported throughputs for the presented design example in \cite{SSP2013} are close to 0.8.
The part of this paper that deals with SIC can be perceived as a generalization of the ideas presented in \cite{SICTAb}, with \JGb{the} important difference that SIC in the proposed framework is performed in the digital domain, effectively removing the memory constraints \cite{PH2010} and potential imperfections of the interference cancellation in the analog domain \cite{ZZ2012}.


%
%
%
%

%
%
\section{Model and Problem Statement} \label{sec:model}

%
%


\PP{We consider a large set of potential transmitters (users) $1,\dots,M$, with $M \gg 1$.} \JG{A small number of users are active and have a message of $D$ bits that should be sent to a common receiver (base station). 
Messages are independent and drawn uniformly at random from $\{1,\dots,2^D\}$.} 
\CSrev{We model the user activity by a single batch arrival, where each user, independently of all other users becomes active (\ie arrives in the batch) with probability $p$ before the protocol that is described below is initiated.
After the protocol has been initiated, we assume that no user arrival occurs; this simplifying assumption allows us to focus on the basic analysis of the protocol operation, similar to other works on tree splitting, c.f. \cite{Massey}. 
In Section~\ref{sec:discussion}, we provide comments on more general models of the user activation and the protocol operation.}

We denote by $\LL$  the set of active users, and by $L=|\LL|$ the number of active users.
Neither the receiver nor the users know $\LL$  or $L$.
Hence, $L$ has a binomial distribution, where the probability that $L$ users are contending is denoted by $q(L) = \binom{M}{L}p^L(1-p)^{M-L}$. For notational convenience, let $q_0=q(0)=(1-p)^M$.
\JG{We denote by $\hat q(L)$  the probability of having $L$ contending users conditioned on the fact there is at least one, \ie} $\hat q(L) = q(L)/(1-q_0)$.

\JGb{The symbol transmitted by the $m-$th user in the $\tau-$th channel use, $\tau\in\mathbb{N}$, is denoted by $X_m(\tau)$.}
\JG{We assume unit channel gains, \ie at} the $\tau-$th channel use the receiver observes 
\begin{equation}\label{eq:PrecodedChannelModel}
Y(\tau) = \sum_{m \in \LL} X_m(\tau) + Z(\tau),
\end{equation}
where $\{Z(\tau)\}_{\tau=1}^\infty$ is white Gaussian noise with unit variance. 

The goal of this paper is to devise \CS{a protocol} that allows the receiver to retrieve both the identities and the messages of all contending users. In particular, we consider \CS{protocol} that \JG{uses multiple blocks} of $N$ channel uses.
In line with other literature on random access, we refer to blocks of channel uses as \emph{slots}. 
\JGb{In each slot the receiver attempts to decode a linear combination of messages transmitted by the users.}
\JG{We restrict our attention to strategies in which the rate (in bits per channel use) is the same for all users and constant over \CS{slots}.} At the end of each slot, the receiver provides feedback to the users and, unless all messages are resolved at the receiver, a new slot is started. Feedback is instantaneous, error free and received by all users. We do not impose any constraints on the amount of feedback that can be provided and explicitly specify how feedback is used later in the paper.

Rephrasing the above, the constituent elements of the protocol are the use of a contention resolution mechanism across slots, dealing with randomness of the user activity pattern, and use of forward error correcting code within slots, dealing with noise. With respect to the latter, we ignore finite block length effects and assume that forward error correcting codes operate with \JGc{vanishing error probability at any rate up to capacity.} As a consequence, the task for the receiver is to recover all packets with \JGc{arbitrarily small} error probability. 


The signal of each user needs to satisfy an average power constraint in each slot, \ie
\begin{equation}
\frac{1}{N}\sum_{\tau=1}^{N} \left|X_m(\tau)\right|^2 \leq P,
\end{equation}
for all $m\in\{1,\dots,M\}$.
We will assume that $P>1$, such that 
a positive computation rate over the multiple access channel can be achieved, as seen in the next section. 
 

%
%
%

We are interested in the following performance parameters.
By $S(L)$ we denote the expected number of slots that the strategy uses to resolve $L$ contending users, where the expectation in $S(L)$ is w.r.t.\ the randomness in the contention resolution mechanism.
By $\Rresolve(L) = L/S(L)$ we denote the expected number of users that is resolved per slot, commonly \JGc{referred} to as throughput in random access literature.
We are also interested in $\barRresolve$, obtained by averaging $\Rresolve(L)$ over $L$, \ie 
\begin{equation}
 \barRresolve = \mathbb{E}[ \Rresolve(L) | L>0 ] = \sum_{L=1}^M \frac{L}{S(L)}\hat q(L).
\end{equation}
In addition to $\barRresolve$, which is obtained under a binomially distributed number of active users as specified above, we also consider the worst-case scenario, \ie we analyze $\Rresolve^*=\inf_L \Rresolve(L)$ under \PP{any model with} \CS{batch arrivals}.

Further, we are interested in the effective number of bits that is transmitted across the channel per channel use (\ie net rate), denoted by $\Rnet(L)$.
Taking into account that $L$ users each transmit $D$ bits in a total of $S(L)$ slots that each consist of $N$ channel uses, we have
\begin{equation}
 \Rnet(L) = \frac{LD}{S(L)N}  = \Rresolve(L)\frac{D}{N}.
\end{equation}
Finally, we are interested in the average and worst-case net rate that we denote by $\barRnet = \mathbb{E}[ \Rnet(L) | L>0 ]$ and $\Rnet^*=\inf_L \Rnet(L)$, respectively.

We will express some of our results in terms of $I_x(a,b)$, the regularized incomplete beta function, which is defined as
\begin{equation} \label{eq:BetaFunc}
I_x(a,b)=B_x(a,b)/B_1(a,b),
\end{equation}
where $B_x(a,b) = \int_0^x t^{a-1}(1-t)^{b-1}dt$. We will use the well-known result that
\begin{equation}
\sum_{L=0}^K q(L) =  I_{1-p}(M-K,K+1) \CS{= 1 - I_p ( K + 1, M - K )}. 
\end{equation}



%
%
\section{Preliminaries} \label{sec:prelim}

%
%

In this section we introduce the three different techniques that constitute the random access mechanism.
In Section~\ref{ssec:plnc}, we present the physical-layer network coding strategy that is adopted in this paper.
Next, we describe a signature coding scheme in Section~\ref{ssec:signaturecodes}.
Finally, we discuss the basic idea of tree splitting in Section~\ref{ssec:treesplitting}.

\subsection{Reliable physical-layer network coding} \label{ssec:plnc}

\PP{The }key ingredient of the random access strategy that is proposed in this paper is to employ physical-layer network coding (PLNC), \ie to organize the physical layer in such a way that the receiver can reliably decode sums of messages that are simultaneously transmitted by users.
This requires a suitable choice of the forward error correcting codes as well as the decoding mechanism that is used by the receiver.
In this section we provide a short introduction to physical-layer network coding and a result from~\cite{nazer11compforw} that will be needed later.
There are various angles at which physical-layer network coding can be approached, \eg denoise-and-forward~\cite{popovski2007physical} or compute-and-forward~\cite{nazer11compforw}.
A survey of these and other approaches is given in~\cite{nazer2011procieee} and~\cite{Liew2013survey}.
In this paper we adopt the compute-and-forward framework, as developed by Nazer and Gastpar in~\cite{nazer11compforw}.

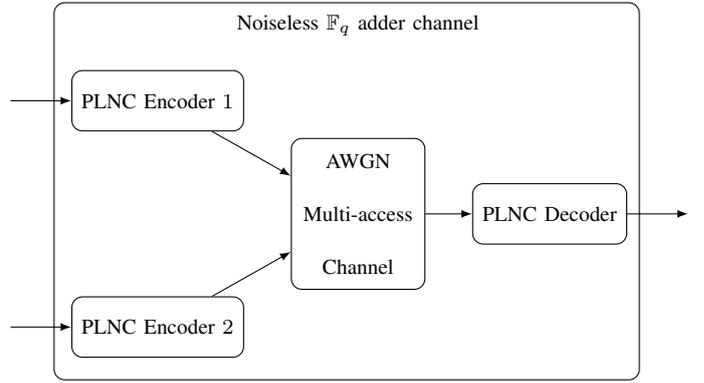
\begin{figure}
\centering
\begin{tikzpicture}
\tikzstyle{boxa}=[draw, rounded corners,minimum width=12mm,minimum height=8mm];
\tikzstyle{boxb}=[draw, circle];
\tikzstyle{->}=[-latex];
\tikzstyle{edgea}=[->];
\tikzstyle{edgeb}=[->,olive];

\node[boxa,anchor=east] (encb) at (-1.5,1.5) {\ft{PLNC Encoder $1$}}; 
\node[boxa,anchor=east] (enca) at (-1.5,-1.5) {\ft{PLNC Encoder $2$}}; 

\node[boxa,minimum height=20mm, text width=15mm, align=center] (channel) at (0,0) {\ft{AWGN Multi-access Channel}};
\draw[edgea] (enca) -- (channel); 
\draw[edgea] (encb) to  (channel); 

\node[boxa,anchor=west] (dec) at (1.5,0) {\ft{PLNC Decoder}};
\draw[edgea] (channel) to (dec); 

\draw[edgea] (dec) -- ($ (dec.east) + (8mm,0) $);
\draw[edgea] ($ (enca.west) - (8mm,0) $) -- (enca);
\draw[edgea] ($ (encb.west) - (8mm,0) $) -- (encb);

\node (addermac) at (0,2.5) {\ft{Noiseless $\FF_q$ adder channel}};
\draw[rounded corners] (-4,-2.2) rectangle (3.7,2.8);

\end{tikzpicture}
\caption{Physical-layer network coding (PLNC) results in a noiseless $\FF_q$ adder channel. ($K=2$ users)} \label{fig:plnc}
\end{figure}

In order to formulate the result from~\cite{nazer11compforw} that we need in the remainder, we consider an arbitrary number of $L$ transmitters.
User $\ell$ has data $W_\ell$ to transmit, where 
\begin{align}
  W_\ell & = ( W_\ell(1), W_\ell(2), \ldots, W_\ell(\kappa) ),
\end{align}
with $W_\ell(j)\in\mathbb{F}_q$, $q$ prime and $\kappa$ is the predefined message length. Each transmitter uses the same linear code $F$ to encode the data into real-valued channel input of length $N$ (\ie the length of a slot) that satisfies an average power constraint $P$. Let $X_\ell = F(W_\ell)$ denote the channel input of user $\ell$. The decoder, upon observing $Y=\sum_{\ell=1}^L X_\ell + Z$ attempts to decode $\sum_\ell W_\ell=(\sum_\ell W_\ell(1), \dots, \sum_\ell W_\ell(\kappa)) $. In this sense, the receiver recovers a function (namely, the sum) of the original messages, which is why this approach is referred to as computation coding. In a sense, as illustrated in Figure~\ref{fig:plnc}, we turn the multi-access AWGN channel in a noiseless $\FF_q$ adder channel.

We denote by $\Rplnc$ the rate of $F$, \ie $\Rplnc = \kappa N^{-1} \log_2 q$ bits per channel use. We will refer to $\Rplnc$ as the computation rate and say that it is achievable if the probability of decoding erroneously can be made arbitrarily small by increasing $N$. The next result follows directly from the main result in~\cite{nazer11compforw}, using the notation
\begin{equation*}
\log^+_2(x) =
\begin{cases}
\log_2(x),\quad &\text{if }x\geq 1, \\
0,\quad &\text{otherwise.} 
\end{cases}
\end{equation*}
\begin{theorem}[\hspace{-4pt} \CS{\cf} \cite{nazer11compforw}, Theorem~1]\label{th:plnc}
For the standard AWGN multiple-access channel, the following computation rate is achievable
\begin{equation}\label{eq:AcheivableComputationRate}
  \Rplnc = \frac{1}{2} \log_2^+ \left( P \right).
\end{equation}
\end{theorem}
The above result does not exactly match the achievable rate as given in\cite[Theorem~1]{nazer11compforw}, which is $\frac{1}{2} \log_2^+ \left( \frac{1}{L} + P \right)$. Since we will be dealing with an unknown number of active users, we use a computation rate that corresponds to $L \rightarrow \infty$ and is thus a lower bound that is valid for arbitrary number of active users.

The proof of Theorem~\ref{th:plnc} in~\cite{nazer11compforw} is based on a random coding argument in which the code $F$ is a lattice that is obtained through Construction A, \cf~\cite{conway2013sphere}.
As a consequence, the value of $q$ is a function of the block length $N$ \JGc{that satisfies $N/q\to 0$ as $N\to\infty$}, see for instance~\cite{erez2004achieving}. \JGc{We will assume that
\begin{equation} \label{eq:qNlogN}
q = N\log_2 N
\end{equation}
in the remainder.}

\subsection{Signature codes} \label{ssec:signaturecodes}

\JGc{We are interested} in signature codes for the $\mathbb{F}_q$ adder channel, when up to $K$ random users, out of total $M$ users, are active. So far, there has been a lot of work investigating the case when the signature symbols are binary, \ie $q=2$; a summary of the known asymptotic results has been presented in \cite{MACbook}. However, the case of general $q$ has been significantly less studied. 
We mention here the construction proposed in \cite{M1990}, which can be generalized to any $q$.
However, this construction results in codewords of length $M$, and allows for simultaneous resolution of up to $K = ( M - 1 ) / 4$ signatures when the set of the contending users is a priori not known.
In this paper we adopt \JG{a} construction that does not require \JG{a} fixed relation of $K$ with respect to $M$ and allows for significantly shorter signatures, as elaborated below.

\subsubsection{A result in additive number theory}
 
In this paper, we adopt the signature code construction presented by Lindstr\"om in~\cite{lindstrom1975}, ~\cite[pp. 42 - 43]{MACbook}. The construction appeared in~\cite{lindstrom1975} for the case $q=2$, but can be readily generalized to arbitrary prime $q$. The construction is designed for the case that the number of users $M$ is a prime; if $M$ is not a prime, one could design signatures for the smallest prime larger than $M$ and use \JGb{only the first} $M$ signatures.
The construction by Lindstr\"om is based on the following result in additive number theory by Bose and Chowla~\cite{bosechowla1962}, which, for convenience of the reader, we present in a slightly less general form than that in~\cite{bosechowla1962}.
\begin{theorem}[\cite{bosechowla1962}] \label{th:bosechowla}
Let $M$ be prime. There exist integers $s_i$, $i=1,\dots,M$, $1\leq s_i< M^K$, such that
\begin{align}
\label{eq:ud}
\sum_{i\in {\cal L}_1} s_i \neq \sum_{i\in {\cal L}_2} s_i,  
\end{align}
for any ${\cal L}_1,{\cal L}_2\subset\{1,\dots,M\}$, $|{\cal L}_1|\leq K$, $|{\cal L}_2|\leq K$ and ${\cal L}_1\neq {\cal L}_2$.
\end{theorem}

Before giving the details, we describe the main idea in Lindstr\"om's construction.
Each integer $s_i$ is expressed through a $r-$ary representation. These $r$-ary representations are mapped to $\mathbb{F}_q$ and used as the signatures of the users.
The choice of $r$ and the mapping to $\mathbb{F}_q$ are such that, from the summation in $\mathbb{F}_q$ of the signatures, the receiver can recover the sum of the integers $s_i$ and thereby the identities of the users.
\JGc{In particular, we will ensure that \JGc{$K(r-1) \leq q-1 $}.} 
Next, we describe the details of our construction.

\subsubsection{Signature encoder}
We assume that a set of integers $s_1,\dots,s_M$, satisfying the conditions of Theorem~\ref{th:bosechowla}, is given. The signature of user $\ell$, denoted by $W_\ell^s$, is a sequence of symbols from $\mathbb{F}_q$.
The signature consists of two parts, each of them with a different functionality.
The first part consists of a single symbol, whose value is fixed to $1$ by each of the users.
In this way, the sum of the first symbols of all active users ${\cal L}$ will provide to the receiver \JGb{$L=|{\cal L}|$}. \JGc{Note that this requires that $M \leq q-1$ and hence a sufficiently long block length, since $q = N\log_2 N$.}
The second part of the signature of user $i$ consists of the $r$-ary representation of $s_i$, where the symbol values are mapped from $\{0,\dots,r-1\}$ to $\mathbb{F}_q$ using the \JGc{natural mapping.}

\subsubsection{Signature decoder}
The signature decoder receives a sum $\sum_{\ell\in {\cal L}} W_\ell^s$, where the set ${\cal L}$ is unknown \JG{to} the receiver.
\CS{The tasks are now the following: i) decide if $L \leq K$, and ii) if $L \leq K$ compute $\LL$; else, treat the received slot as a collision slot.}

The first operation of the decoder is to map the sequence $\sum_{\ell\in {\cal L}} W_\ell^s$ from $\mathbb{F}_q$ to a sequence of integers using a natural mapping. 
Through the sum of the first symbols, the receiver detects the number of active users $L$. 

Next, consider the case when $L \leq K$.
Since \JGc{$K(r-1) \leq q-1$}, the addition of at most $K$ symbols $\{0,\dots,r-1\}$ in $\mathbb{F}_q$ will be equivalent to the addition over the integers.
Therefore, the decoder can recover the sum of the $r$-ary representations of the $s_\ell$, $\ell\in\mathcal{L}$.
More precisely, each symbol in this summation is an element from $\{0,\dots,K(r-1)\}$ that is obtained by adding integers from $\{0,\dots,r-1\}$.
The resulting sum immediately provides $\sum_{i\in {\cal L}} s_i$ and, by Theorem~\ref{th:bosechowla}, also provides ${\cal L}$.


In case a collision is detected and $L > K$, the receiver can reliably compute the sum of the data of the active users, but cannot \JGb{decode} the identities of the active users, \ie who contributed to the sum.
The resolution of the active user identities and their data packets in this case is \JGb{addressed by the proposed scheme, as introduced in} Section~\ref{sec:idea}.

\subsubsection{Analysis of signature length}

\JGc{
We now fix the value of $r$ to
\begin{equation}
r = \left\lfloor \frac{q-1}{K} \right\rfloor,
\end{equation}
such that $K(r-1) \leq q-1$ is satisfied.} \JGb{This leads to the following upper bound on the lengths of the signatures.} \JGc{We express the length in the number of $q$-ary symbols that are used. The reason is that this number will determine the overhead of the signatures in the overall scheme as introduced in Section~\ref{sec:idea}.}
\begin{theorem} \label{th:sigcode}
There exist signatures of length
\JGc{
\begin{equation}
N_w = \left\lceil\frac{K\log_2 M}{\log_2(q-1) - \log_2 K}\right\rceil + 1, 
\end{equation}
$r$-ary symbols, where $r = \left\lfloor (q-1)/K \right\rfloor.$ These $r$-ary symbols can be represented by an equal number of $q$-ary symbols.
}
\end{theorem}
\begin{IEEEproof}
The length of \JGc{the signatures} follows from Theorem~\ref{th:bosechowla}. In particular, the signatures are represented by $K\log_r M + 1$ long strings of \JGc{$r$-ary symbols. The representation in $q$-ary symbols follows because $r\leq q$.}
\end{IEEEproof}
\JGc{In the proposed random access strategy we will concatenate our signatures with a message. We will achieve reliable communication by considering asymptotically long messages, \ie we let $N\to\infty$. This implies that $q\to\infty$, since $q=N\log_2 N$. The influence of the signatures on the rate is negligible, \ie the ratio $N_w/N$ is vanishing.}

\subsection{Tree splitting} \label{ssec:treesplitting}
We briefly outline the basic binary tree-splitting algorithm under a collision model~\cite{capetanakis}.
Again, $\LL$ denotes the set of active users \JGb{and} $L=|\LL|$, $1\leq L\leq M$, denotes the number of active users.
In the first slot all $L$ users transmit a packet.
If $L = 1$, the receiver successfully decodes the packet of the user and the contention period ends.
If $L\geq 2$, a collision occurs and \JGc{the receiver cannot decode the packets.} 
The users then probabilistically split into two groups $\LL_{1}$ and $\LL_{2}$.
The splitting is uniform at random and independent over users, \ie each user flips a fair coin to decide on the group to join.
Both groups then contend for the medium in the same fashion: first the users from $\LL_1$, then the users from $\LL_2$.
The splitting is done recursively, eventually leading to an instance in which only a single user is active and the corresponding transmission is successfully received.
The algorithm continues until the transmissions of all active users from $\LL$ are successfully received.
By means of feedback, after each slot the receiver informs the users whether there was a collision, a single \JGb{transmission}, or no transmission present, directing the future actions of the active users.

The above described algorithm and its variations were thoroughly analyzed in the literature, assessing the performance parameters such as throughput, delay and stability.
The work closest to ours is presented in~\cite{Massey}, the most important difference being that we investigate a generalized case in which collisions occur when $L > K$, where $K \geq 1$.
The related analysis, which also covers the special case $K=1$, is presented in Section V.


%
%
%
\section{The Proposed Strategy} \label{sec:idea}

\begin{figure}
\centering
\begin{tikzpicture}[scale=.45]
\tikzstyle{H}=[draw, rectangle,minimum width=6mm,minimum height=6mm];
\tikzstyle{boxb}=[draw, circle,inner sep=1pt];
\tikzstyle{boxc}=[fill=white, circle,minimum width=5mm];
\tikzstyle{->}=[-latex];

\node (ell) at (-2,1) {\fm{\ell}};

\node[draw,rounded corners] (sigcode) at (-2,-2) {\ft{Signature encoder}};

\begin{scope}[xshift=0cm,yshift=1cm]
\draw (0,-.6) rectangle (6,.6);
\foreach \x in {1, 2, ..., 5}
 {
    \draw (\x,-.6) -- (\x,.6);
}
\node[fill=white,inner sep=1pt,minimum width=20mm] at (3,0) {\ft{Data:} \fm{W_\ell^d}};
\end{scope}

\begin{scope}[xshift=-4cm,yshift=-5cm]
\draw (0,-.6) rectangle (10,.6);
\foreach \x in {1, 2, ..., 9}
 {
    \draw (\x,-.6) -- (\x,.6);
}
\draw[ultra thick] (4,-.6) -- (4,.6);
\node[fill=white,inner sep=1pt,minimum width=10mm] at (2,0) {\fm{W_\ell^s}};
\node[fill=white,inner sep=1pt,minimum width=20mm] at (7,0) {\fm{W_\ell^d}};
\end{scope}

\node[draw,rounded corners] (plnc) at (10,-5) {\ft{PLNC encoder}};

\begin{scope}[xshift=3cm,yshift=-10cm]
\draw (0,-.6) rectangle (10,.6);
\foreach \x in {1, 2, ..., 9}
 {
    \draw (\x,-.6) -- (\x,.6);
}
\node[fill=white,inner sep=1pt,minimum width=40mm] at (5,0) {\fm{X_\ell}};
\end{scope}

\draw[rounded corners,dashed] (-5,-1) rectangle (13,-8);
\node at (3.3,-7) {\ft{Encoder}};

\draw[->] (ell) -- (sigcode);
\draw[->] (-2,-2.5) -- (-2,.-4.4);
\draw[->] (3,.4) -- (3,-4.4);
\draw[->] (6,-5) -- (plnc);
\draw[->] (plnc) -- (10,-9.5);

\end{tikzpicture}
\caption{Illustration of the encoder for user $\ell$ in a slot.\label{fig:encoder}}
\end{figure}
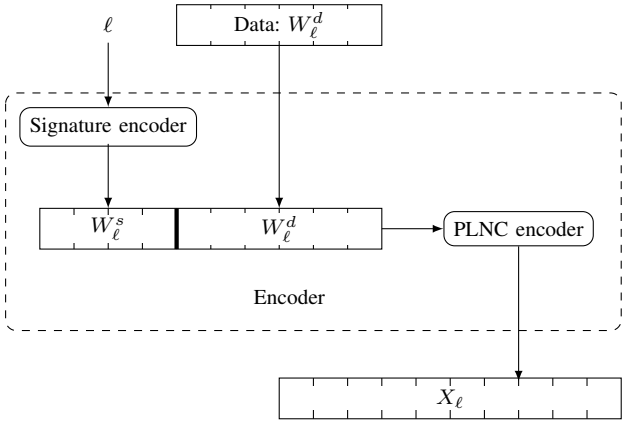

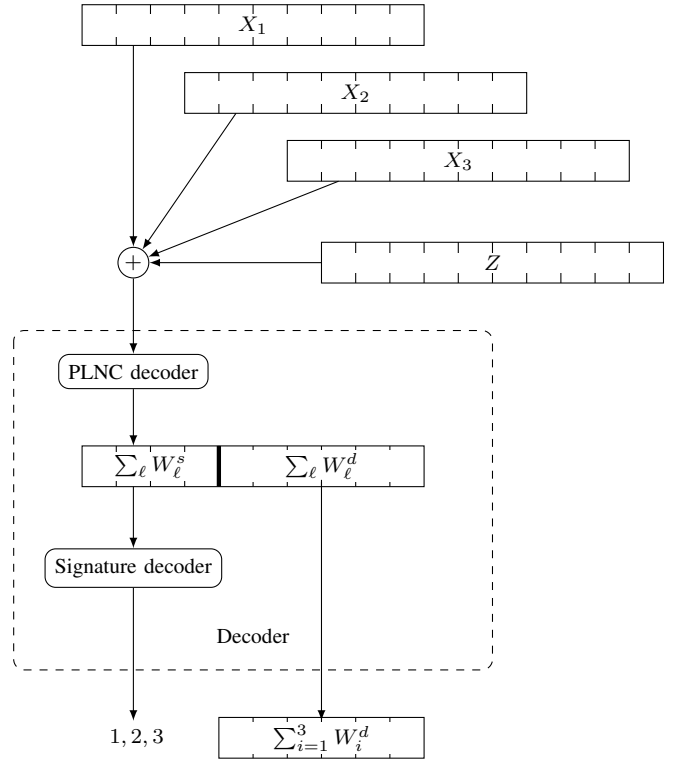
\begin{figure}
\centering
\begin{tikzpicture}[scale=.45]
\tikzstyle{H}=[draw, rectangle,minimum width=6mm,minimum height=6mm];
\tikzstyle{boxb}=[draw, circle,inner sep=1pt];
\tikzstyle{boxc}=[fill=white, circle,minimum width=5mm];
\tikzstyle{->}=[-latex];

\begin{scope}[xshift=0cm,yshift=7cm]
\draw (0,-.6) rectangle (10,.6);
\foreach \x in {1, 2, ..., 9}
 {
    \draw (\x,-.6) -- (\x,.6);
}
\node[fill=white,inner sep=1pt,minimum width=40mm] at (5,0) {\fm{X_{1}}};
\end{scope}

\begin{scope}[xshift=3cm,yshift=5cm]
\draw (0,-.6) rectangle (10,.6);
\foreach \x in {1, 2, ..., 9}
 {
    \draw (\x,-.6) -- (\x,.6);
}
\node[fill=white,inner sep=1pt,minimum width=40mm] at (5,0) {\fm{X_{2}}};
\end{scope}

\begin{scope}[xshift=6cm,yshift=3cm]
\draw (0,-.6) rectangle (10,.6);
\foreach \x in {1, 2, ..., 9}
 {
    \draw (\x,-.6) -- (\x,.6);
}
\node[fill=white,inner sep=1pt,minimum width=40mm] at (5,0) {\fm{X_{3}}};
\end{scope}

\begin{scope}[xshift=7cm,yshift=0cm]
\draw (0,-.6) rectangle (10,.6);
\foreach \x in {1, 2, ..., 9}
 {
    \draw (\x,-.6) -- (\x,.6);
}
\node[fill=white,inner sep=1pt,minimum width=40mm] at (5,0) {\fm{Z}};
\end{scope}

\node[boxb]  (pl) at (1.5,0) {\fm{+}};

\node[draw,rounded corners,anchor=north] (plnc) [below=10mm of pl] {\ft{PLNC decoder}};

\begin{scope}[xshift=0cm,yshift=-6cm]
\draw (0,-.6) rectangle (10,.6);
\foreach \x in {1, 2, ..., 9}
 {
    \draw (\x,-.6) -- (\x,.6);
}
\draw[ultra thick] (4,-.6) -- (4,.6);
\node[fill=white,inner sep=1pt,minimum width=10mm] at (2,0) {\fm{\sum_\ell W^s_\ell}};
\node[fill=white,inner sep=1pt,minimum width=20mm] at (7,0) {\fm{\sum_\ell W^d_\ell}}; 
\end{scope}

\node[draw,rounded corners] (sigcode) at (1.5,-9) {\ft{Signature decoder}};

\begin{scope}[xshift=4cm,yshift=-14cm]
\draw (0,-.6) rectangle (6,.6);
\foreach \x in {1, 2, ..., 5}
 {
    \draw (\x,-.6) -- (\x,.6);
}
\node[fill=white,inner sep=1pt,minimum width=20mm] at (3,0) {\fm{\sum_{i=1}^3 W^d_{i}}};
\end{scope}

\node[anchor=west] (ell) at (.5,-14) {\fm{1,2,3}};

\draw[rounded corners,dashed] (-2,-2) rectangle (12,-12);
\node at (5,-11) {\ft{Decoder}};

\draw[->] (1.5,6.4) -- (pl);
\draw[->] (4.5,4.4) -- (pl);
\draw[->] (7.5,2.4) -- (pl);
\draw[->] (7,0) -- (pl);
\draw[->] (pl) -- (plnc);
\draw[->] (plnc) -- (1.5,-5.4);
\draw[->] (7,-6.4) -- (7,-13.5);
\draw[->] (1.5,-6.6) -- (sigcode);
\draw[->] (sigcode) -- (1.5,-13.5);

\end{tikzpicture}
\caption{Illustration of the decoder in a slot. ($L=3$ users). \label{fig:decoder}}
\end{figure}

%
%


We start with an overview of the proposed random access strategy.
The strategy operates in rounds, \CS{where a round includes (i) a slot in which the active users transmit the PLNC encoded concatenation of their signatures and payloads, and (ii) the corresponding feedback from the common receiver}.
The use of PLNC enables the receiver to reliably obtain the $q-$ary sums of the user transmissions.
As long as there are at most $K$ active users, the receiver is able to uniquely decode their signatures, detect which users are active and direct them towards solving the linear combination of their payloads.

The receiver is also able to detect when more than $K$ users \JGb{are} active \CS{via the sum of indicator symbols contained in the signatures}.
In this case, the receiver instructs the users to randomly split in two groups and the strategy is then executed in a recursive fashion for each of these groups.
We proceed by presentation of the details.

\subsection{Encoder} \label{ssec:ideaencoder}

Let $W_\ell^s$ and $W_\ell^d$ denote the strings representing the signature and the data payload, respectively, of the active user $\ell$.
The concatenation of signature and payload $W_\ell = W_\ell^s \| W_\ell^d$ is used as the input of a PLNC encoder.
Recall from Section~\ref{ssec:plnc} that the PLNC encoder applies a linear forward error correcting code, the same code $F$ for all users.
The output of the PLNC encoder, denoted by $X_\ell = F(W_\ell) = F(W_\ell^s \| W_\ell^d)$, is a channel input of user $\ell$. The operation of the encoder of a single user in a \JGb{slot} is illustrated in Figure~\ref{fig:encoder}.

\subsection{Decoder} \label{ssec:ideadecoder}

The receiver observes $Y$, which is a \emph{real} sum  of $X_\ell$, $\ell\in\LL$, and additive noise $Z$,
\begin{equation}
Y =  \sum_{\ell\in\LL} X_\ell + Z =  \sum_{\ell\in\LL} F(W_\ell) + Z.
\end{equation}
It uses a PLNC decoder to decode $Y$ and obtain
\begin{equation}
\sum_{\ell\in\LL} W_\ell, 
\end{equation}
which decomposes into the sums of the signatures $\sum_{\ell\in\LL} W^s_\ell$ and the sums of the codewords $\sum_{\ell\in\LL} W_\ell^d$. This is illustrated in Figure~\ref{fig:decoder}.
\JGb{As explained in Section~\ref{ssec:plnc}, since the first symbol in the signatures of all users is $1$, the receiver directly obtains the number of active users $L=|\LL|$, }

\subsection{User resolution for $L\leq K$}

\JGb{It \JGc{follows} from the properties of the signature code, that if} $L \leq K$,  the receiver obtains \JGb{$\LL$.}
Moreover, it also has the sum of the messages $\sum_{\ell\in\LL} W_\ell^d$.
By making use of the feedback mechanism to the users, the receiver ensures that in the next $L-1$ rounds $L-1$ of the users in $\LL$ are individually transmitting their messages.
This can be achieved by, \eg polling the users through the feedback at the end of a round \PP{and requesting them to transmit in the next \CS{slot}}. In that case the feedback acts as an ACK as well as a scheduling mechanism. \JG{Note that such a polling mechanism is not possible in the case when $L>K$, since the receiver then does not know the identities of the users.}

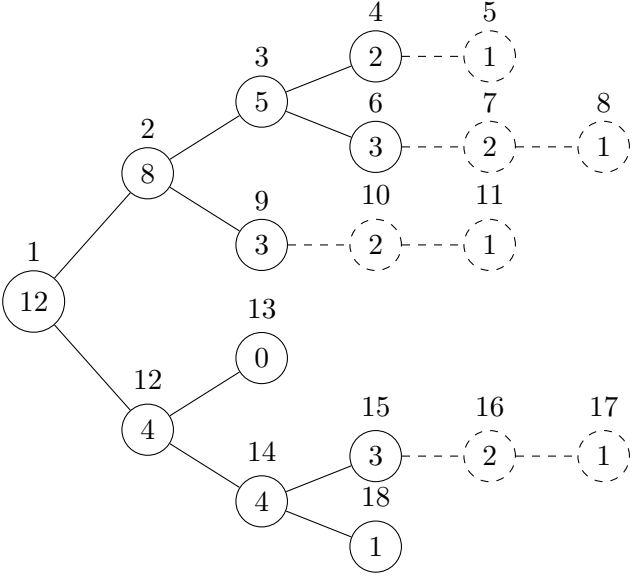
\begin{figure}
\centering
\begin{tikzpicture}[scale=1]
\path[
  every node/.style={circle,draw, label distance=-1mm},
  level distance=15mm, 
  level 1/.style={sibling distance=34mm},
  level 2/.style={sibling distance=19mm},
  level 3/.style={sibling distance=12mm},
]
  node[label=$1$] {$12$}[grow'=right]
    child {node[label=$2$] {$8$}
      child {node[label=$3$] {$5$}
        child {node[label=$4$] {$2$}
          child [dashed] {node[label=$5$] {$1$}
          }
        }
        child {node[label=$6$] {$3$}
          child [dashed] {node[label=$7$] {$2$}
            child {node[label=$8$] {$1$}
            }
          }
        }
      }
      child {node[label=$9$] {$3$}
        child [dashed] {node[label=$10$] {$2$}
          child {node[label=$11$] {$1$}
          }
        }
      }
    }
    child {node[label=$12$] {$4$}
      child {node[label=$13$] {$0$}
      }
      child {node[label=$14$] {$4$}
        child  {node[label=$15$] {$3$}
          child [dashed] {node[label=$16$] {$2$}
            child {node[label=$17$] {$1$}
            }
          }
        }
        child {node[label=$18$] {$1$}
        }
      }
    };
\end{tikzpicture}
\caption{Illustration of tree splitting, $K=3$. Each node represents a transmission. The label above a node indicates the slot in which the transmission takes place and the number inside a node indicates the number of transmitting users. The dashed nodes (and corresponding edges) are obtained through polling/scheduling instead of splitting.}
\label{fig:tree}
\end{figure}

\subsection{User resolution for $L>K$} \label{ssec:ideasplit}

When the receiver observes $L > K$, then this is a ``collision'' in our generalized setting and the receiver signals this fact via feedback.
All users in $\LL$ now participate in a splitting protocol with uniform splits into two groups.
Each user independently of the other users draws a uniformly distributed random number from $\{1,2\}$.
All users with value $1$ enter a new contention resolution phase.
The users with value $2$ wait until this phase ends and start another contention resolution phase afterwards.
If there are more than $K$ users in one of these groups the splitting procedure is applied recursively.
The splitting protocol is illustrated in Figure~\ref{fig:tree}.

In the next section we analyze the proposed strategy, \JGc{parameterized} on the values of $K$.
Note that case $K =1$ reduces the scheme to the traditional tree splitting protocol that was discussed in Section~\ref{ssec:treesplitting}.

\section{Analysis} \label{sec:analysis}

%
%

\subsection{Expected length of the contention resolution phase and expected throughput}

We provide an analysis in terms of a recursive expression for the expected number of slots in a contention resolution period \CS{given} the number of active users $L$, denoted as $S(L)$.
The analysis is similar to the one by Massey~\cite{Massey} that deals with the case $K=1$.
Let
\begin{align}
\alpha^* &= 1+\frac{1}{K}, \\
\beta^* &= 1 + \frac{2}{(K+1)(1-2^{-K})}. \label{eq:beta}
\end{align}
We will show that
\begin{equation}
\alpha^*L-1 \leq S(L) \leq \beta^*L-1.
\end{equation}

Let $p_L(\ell)$ denote the probability that a group of $L$ users split into two groups where one of the groups has size $\ell$.
We have 
\begin{equation}
p_L(\ell) = \binom{L}{\ell}2^{-L}.
\end{equation}
Note that $p_L(0)>0$, \ie it is possible that there are groups with no users, thus
we also include the case that $L=0$.
We start by stating the recursion:
\begin{lemma} \label{lem:recursion}
\begin{equation} \label{eq:SLrecursion}
S(L) =
\begin{cases}
1,\quad &\text{if } L=0,\\
L,\quad &\text{if } 1\leq L\leq K,\\
\displaystyle\frac{1+ 2\sum_{i=0}^{L-1}p_L(i)S(i)}{1-2p_L(L)},\quad &\text{if } L>K.
\end{cases}
\end{equation}
\end{lemma}
\begin{IEEEproof}
Since we have a $K$ out of $M$ signature code, we have $S(0)=1$, $S(1)=1$, $S(2)=2,\dots,S(K)=K$. For $L>K$ we have the following recursion
\begin{align} 
S(L)
&= 1 + \sum_{i=0}^L p_L(i)\left\{S(i)+S(L-i)\right\} \label{eq:Snosic} \\
&= 1 + \sum_{i=0}^L \left\{p_L(i)S(i)+p_L(L-i)S(L-i)\right\} \\
&= 1 + 2\sum_{i=0}^Lp_L(i)S(i),
\end{align}
which can be rewritten as
\begin{equation}
S(L) = \frac{1+ 2\sum_{i=0}^{L-1}p_L(i)S(i)}{1-2p_L(L)},
\end{equation}
by making use of $\binom{L}{L-i}=\binom{L}{i}$ and $\sum_{i=1}^Lp_L(i)S(L-i)=\sum_{i=1}^Lp_L(i) S(i)$.
\end{IEEEproof}

For notational convenience, let $\gamma(L)$, $L>K$, be defined as 
\begin{equation} \label{eq:gammadef}
\gamma(L) = \frac{\sum_{i=0}^K \left( S(i) + 1\right)p_L(i)}{\sum_{i=0}^K p_L(i) i} = 1 + \frac{1 + \sum_{i=0}^K \binom{L}{i}}{\sum_{i=0}^K\binom{L}{i}i}.
\end{equation}
The reason for introducing $\gamma(L)$ is that it can be used to express bounds on $S(L)$, as demonstrated next. 
\begin{lemma} \label{lem:TLbound}
If $\alpha$ and $\beta$ satisfy
\begin{equation} \label{eq:lemboundconditons}
\alpha \leq \gamma(L) \leq \beta
\end{equation}
for all $L>K$, then
\begin{equation} \label{eq:lembounds}
\alpha L - 1 \leq S(L) \leq \beta L - 1
\end{equation}
for all $L>K$.
\end{lemma}
 \begin{IEEEproof}
For all $L\leq K$ we have
\begin{equation} \label{eq:prbetabound1}
S(L) \leq \beta L - 1 +\sum_{i=0}^K \delta_{iL}(S(L)-\beta L +1),
\end{equation}
where $\delta_{ij}$ is the Kronecker delta, defined to be $1$ if $i=j$ and $0$ otherwise.
We now induction on $L$ to show that~\eqref{eq:prbetabound1} holds for all $L$, thereby providing the proof for the upper bound in~\eqref{eq:lembounds}.

As an induction hypothesis, assume that~\eqref{eq:prbetabound1} holds for $S(K + 1), S( K + 2), \dots, S(L-1)$, $L > K + 1$.
Then, we have the following bound for $S(L)$ 
\begin{align}
S(L)
=&\ \frac{1 + 2\sum_{i=0}^{L-1} p_L(i)S(i)}{1-2p_L(L)} \label{eq:prbetaboundind1} \\
\leq&\ \frac{1 + 2\sum_{i=0}^{L-1} p_L(i)(\beta i -1)}{1-2p_L(L)}  \notag \\
&\ + \frac{2 \sum_{i=0}^K p_L(i)  (S(i) -\beta i + 1) }{1-2p_L(L)} \label{eq:prbetaboundind2} \\
= & \ \frac{1 + 2\sum_{i=0}^{L-1} p_L(i)(\beta i -1)}{1-2p_L(L)}  \notag \\
&\ + \frac{2 ( \gamma ( L ) - \beta ) \sum_{i=0}^K p_L(i)  i }{1-2p_L(L)} \label{eq:prbetaboundind2a} \\
\leq&\ \frac{1 + 2\sum_{i=0}^{L-1} p_L(i)(\beta i -1)}{1-2p_L(L)} \label{eq:prbetaboundind3} \\
=&\ \beta L -1,  \label{eq:prbetaboundind4}
\end{align}
where~\eqref{eq:prbetaboundind1} follows from Lemma~\ref{lem:recursion}, \eqref{eq:prbetaboundind2} from the induction hypothesis, \eqref{eq:prbetaboundind2a} from the definition of $\gamma(L)$ in~\eqref{eq:gammadef}, \eqref{eq:prbetaboundind3} from condition~\eqref{eq:lemboundconditons}, and finally,  \eqref{eq:prbetaboundind4} from
\begin{align}
& \sum_{i=0}^{L-1} p_L(i) i = \sum_{i=0}^{L} p_L(i) i - p_L(L)L = \frac{L}{2}(1-2p_L(L)), \\
& \sum_{i=0}^{L-1} p_L(i) = 1 - p_L ( L ).
\end{align}
The establishes the upper bound on $S(L)$. The proof of the lower bound follows in entirely analogous fashion. 
 \end{IEEEproof}

Next, we provide an upper and a lower bound on $\gamma(L)$. 
\begin{lemma} \label{lem:gammabound}
\begin{equation}
1+\frac{1}{K} \leq \gamma(L) \leq 1 + \frac{2}{(K+1)(1-2^{-K})}.
\end{equation}
\end{lemma}
\begin{IEEEproof}
For the lower bound we have
\begin{align}
\gamma(L)
&= 1 + \frac{1 + \sum_{i=0}^K \binom{L}{i}}{\sum_{i=0}^K\binom{L}{i}i} \\
&\geq 1 + \frac{\sum_{i=0}^K\binom{L}{i} }{K\sum_{i=0}^K\binom{L}{i}} \\
&= 1 + \frac{1}{K} = \alpha^{*}.
\end{align}
In order to prove the upper bound, we first show that $\gamma(L)$ is decreasing in $L$, \ie that $\gamma(L)-\gamma(L+1)\geq 0$.
It immediately follows from~\eqref{eq:gammadef} that $\gamma(L)-\gamma(L+1)\geq 0$ if
\begin{multline} \label{eq:gammabound0}
\sum_{i=0}^K\binom{L+1}{i}i 
+ \sum_{i=0}^K\binom{L}{i}\sum_{j=0}^K\binom{L+1}{j}j \\
- \sum_{i=0}^K\binom{L}{i}i - \sum_{i=0}^K\binom{L+1}{i}\sum_{j=0}^K\binom{L}{j}j \geq 0.
\end{multline}
Since $\binom{L+1}{i}  \geq \binom{L}{i} $ for all $0\leq i\leq K<L$, it is sufficient to show that
\begin{equation} \label{eq:gammabound1}
\sum_{i=0}^K\binom{L}{i}\sum_{j=0}^K\binom{L+1}{j}j - \sum_{i=0}^K\binom{L+1}{i}\sum_{j=0}^K\binom{L}{j}j
\end{equation}
is non-negative. Using $\binom{L+1}{i}=\frac{L+1}{L+1-i}\binom{L}{i}$ we rewrite~\eqref{eq:gammabound1} as
\begin{equation} \label{eq:gammabound2}
\sum_{i,j=0}^K\left( \frac{L+1}{L+1-j} - \frac{L+1}{L+1-i}\right) j \binom{L}{i}\binom{L}{j}.
\end{equation}
Now, \eqref{eq:gammabound0} follows from the observation that negative terms occur in~\eqref{eq:gammabound2} for $i>j$, in which case, by symmetry considerations, there are corresponding positive  $j > i $ terms that are $ j - i $ times larger by absolute values than their negative counterparts.
The upper bound now follows directly from the value of $\gamma(L)$ at $L=K+1$, \ie
 \begin{align}
 \gamma (K+1)
 &= 1 + \frac{1 + \sum_{i=0}^K\binom{K+1}{i} }{\sum_{i=0}^K\binom{K+1}{i}i} \\
 &= 1 + \frac{1 + (2^{K+1}-1)}{2^K(K+1) - (K+1)} \\ 
 & = 1 + \frac{ 2 }{ ( K+1 )( 1 - 2^{ - K })} = \beta^{*}.
 \end{align}
\end{IEEEproof}

We are now ready to state the main result of this section.
\begin{theorem} \label{th:boundsbasic}
$S(L) = L $ if $1\leq L\leq K$, and, for $L>K$
\begin{equation}
\alpha^* L - 1 \leq S(L) \leq \beta^* L - 1.
\end{equation}
\end{theorem}
\begin{IEEEproof}
Directly from Lemmas~\ref{lem:TLbound} and~\ref{lem:gammabound}.
\end{IEEEproof}

%
%
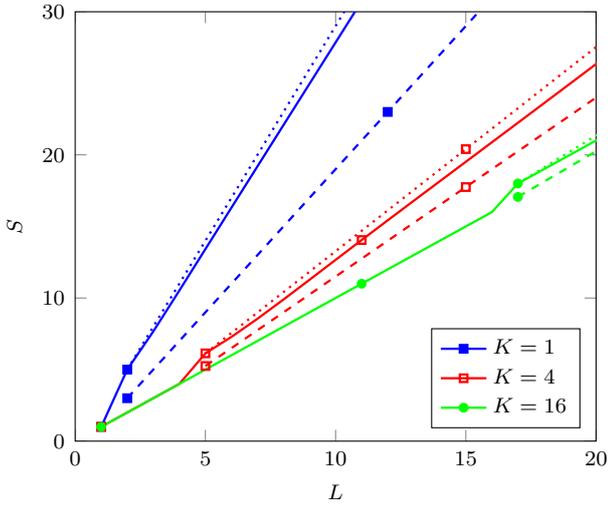
\begin{figure}
\centering
\begin{tikzpicture}
\begin{axis}[
  xlabel=$L$,ylabel=$S$, 
  xmin=0, xmax=20,
  ymin=0,ymax=30,
  font=\scriptsize,
  legend style={
        cells={anchor=west},
        legend pos=south east,
       font=\scriptsize,
    }
]

\addplot[
  line width=.3mm,color=blue, solid,
  mark=square*,mark repeat=10,mark phase=0,mark size=.5mm,mark options={solid}
  ]
table[
  header=false,x index=0,y index=1,
  ]
{matlab_figures/SinL.csv};
\addlegendentry{$K=1$};
\addplot[
  line width=.3mm,color=blue, dashed,
 mark=square*,mark repeat=10,mark phase=0,mark size=.5mm,mark options={solid},
 forget plot
  ]
table[
  header=false,x index=0,y index=2,
  ]
{matlab_figures/SinL.csv};
\addplot[
  line width=.3mm,color=blue, dotted,
 mark=square*,mark repeat=10,mark phase=0,mark size=.5mm,mark options={solid},
 forget plot
  ]
table[
  header=false,x index=0,y index=3,
  ]
{matlab_figures/SinL.csv};
\addplot[
  line width=.3mm,color=red, solid,
  mark=square,mark repeat=10,mark phase=0,mark size=.5mm,mark options={solid}
  ]
table[
  header=false,x index=0,y index=4,
  ]
{matlab_figures/SinL.csv};
\addlegendentry{$K=4$};
\addplot[
  line width=.3mm,color=red, dashed,
  mark=square,mark repeat=10,mark phase=0,mark size=.5mm,mark options={solid},
forget plot
  ]
table[
  header=false,x index=0,y index=5,
  ]
{matlab_figures/SinL.csv};
\addplot[
  line width=.3mm,color=red, dotted,
  mark=square,mark repeat=10,mark phase=0,mark size=.5mm,mark options={solid},
forget plot
  ]
table[
  header=false,x index=0,y index=6,
  ]
{matlab_figures/SinL.csv};
\addplot[
  line width=.3mm,color=green, solid,
  mark=*,mark repeat=10,mark phase=0,mark size=.5mm,mark options={solid}
  ]
table[
  header=false,x index=0,y index=7,
  ]
{matlab_figures/SinL.csv};
\addlegendentry{$K=16$};
\addplot[
  line width=.3mm,color=green, dashed,
  mark=*,mark repeat=10,mark phase=0,mark size=.5mm,mark options={solid},
 forget plot
  ]
table[
  header=false,x index=0,y index=8,
  ]
{matlab_figures/SinL.csv};
\addplot[
  line width=.3mm,color=green, dotted,
  mark=*,mark repeat=10,mark phase=0,mark size=.5mm,mark options={solid},
 forget plot
  ]
table[
  header=false,x index=0,y index=9,
  ]
{matlab_figures/SinL.csv};
\end{axis}
\end{tikzpicture}
\caption{$S(L)$ and its bounds for various values of $K$. Upper and lower bounds in dotted and dashed lines, respectively. Exact values of $S(L)$ in solid lines.\label{fig:SinL}}
\end{figure}

%
%
\begin{figure}
\centering
\begin{tikzpicture}
\begin{axis}[
  xlabel=$L$,ylabel=$\Rresolve$, 
  xmin=0, xmax=20,
  ymin=0,ymax=1.2,
  font=\scriptsize,
  legend style={
        cells={anchor=west},
        legend pos=south east,
       font=\scriptsize,
    }
]

\addplot[
  line width=.3mm,color=blue, solid,
  mark=square*,mark repeat=10,mark phase=0,mark size=.5mm,mark options={solid}
  ]
table[
  header=false,x index=0,y index=1,
  ]
{matlab_figures/RresinL.csv};
\addlegendentry{$K=1$};
\addplot[
  line width=.3mm,color=blue, dashed,
 mark=square*,mark repeat=10,mark phase=0,mark size=.5mm,mark options={solid},
 forget plot
  ]
table[
  header=false,x index=0,y index=2,
  ]
{matlab_figures/RresinL.csv};
\addplot[
  line width=.3mm,color=blue, dotted,
 mark=square*,mark repeat=10,mark phase=0,mark size=.5mm,mark options={solid},
 forget plot
  ]
table[
  header=false,x index=0,y index=3,
  ]
{matlab_figures/RresinL.csv};
\addplot[
  line width=.3mm,color=red, solid,
  mark=square,mark repeat=10,mark phase=0,mark size=.5mm,mark options={solid}
  ]
table[
  header=false,x index=0,y index=4,
  ]
{matlab_figures/RresinL.csv};
\addlegendentry{$K=4$};
\addplot[
  line width=.3mm,color=red, dashed,
  mark=square,mark repeat=10,mark phase=0,mark size=.5mm,mark options={solid},
forget plot
  ]
table[
  header=false,x index=0,y index=5,
  ]
{matlab_figures/RresinL.csv};
\addplot[
  line width=.3mm,color=red, dotted,
  mark=square,mark repeat=10,mark phase=0,mark size=.5mm,mark options={solid},
forget plot
  ]
table[
  header=false,x index=0,y index=6,
  ]
{matlab_figures/RresinL.csv};
\addplot[
  line width=.3mm,color=green, solid,
  mark=*,mark repeat=10,mark phase=0,mark size=.5mm,mark options={solid}
  ]
table[
  header=false,x index=0,y index=7,
  ]
{matlab_figures/RresinL.csv};
\addlegendentry{$K=16$};
\addplot[
  line width=.3mm,color=green, dashed,
  mark=*,mark repeat=10,mark phase=0,mark size=.5mm,mark options={solid},
 forget plot
  ]
table[
  header=false,x index=0,y index=8,
  ]
{matlab_figures/RresinL.csv};
\addplot[
  line width=.3mm,color=green, dotted,
  mark=*,mark repeat=10,mark phase=0,mark size=.5mm,mark options={solid},
 forget plot
  ]
table[
  header=false,x index=0,y index=9,
  ]
{matlab_figures/RresinL.csv};
\end{axis}
\end{tikzpicture}
\caption{$\Rresolve(L)$ and its bounds for various values of $K$. Upper and lower bounds in dotted and dashed lines, respectively. Exact values of $\Rresolve(L)$ in solid lines.\label{fig:RresinL}}
\end{figure}

\begin{table}
\centering
\begin{equation*}
\begin{array}{rll}
K & \alpha^* & \beta^* \\
\hline
1 &  2         & 3 \\
2 &  1.5      & 1.889\\
4 &  1.25    & 1.427 \\
8 &  1.125  & 1.223 \\
16 & 1.063 & 1.118
\end{array}
\end{equation*}
\caption{Values for $\alpha^*$ and $\beta^*$ that serve in the bounds on $S(L)$.} \label{table:TLbounds}
\end{table}

In Table~\ref{table:TLbounds} we provide a numerical evaluation of $\alpha^*$ and $\beta^*$.
Also, in Figure~\ref{fig:SinL} we illustrate $S(L)$, as well as its lower and upper bounds for various values of $K$. Finally, in Figure~\ref{fig:RresinL} we illustrate the expected number of users that is resolved per slot given that $L$ users are active, $\Rresolve(L) = L/S(L)$, including its upper and lower bounds derived from the bounds on $S( L )$.

\begin{figure}
\centering
\begin{tikzpicture}
\begin{axis}[
  xlabel=$K$,ylabel=$\barRresolve$, 
  font=\scriptsize,
  legend style={
        cells={anchor=west},
        legend pos=south east,
       font=\scriptsize,
    }
]

\addplot[
  line width=.3mm,color=blue, solid,
  mark=square*,mark size=.5mm,mark options={solid}
  ]
table[
  header=false,x index=0,y index=1,
  ]
{matlab_figures/avgperf_inK.csv};
\addlegendentry{$pM=3$};
\addplot[
  line width=.3mm,color=red, solid,
  mark=square,mark size=.5mm,mark options={solid}
  ]
table[
  header=false,x index=0,y index=2,
  ]
{matlab_figures/avgperf_inK.csv};
\addlegendentry{$pM=6$};
\addplot[
  line width=.3mm,color=green, solid,
  mark=*,mark size=.5mm,mark options={solid}
  ]
table[
  header=false,x index=0,y index=3,
  ]
{matlab_figures/avgperf_inK.csv};
\addlegendentry{$pM=12$};
\end{axis}
\end{tikzpicture}
\caption{Lower bounds on $\barRresolve$, the expected number of users that is resolved per slot.  ($M=1031$)\label{fig:avgperf_inK}}
\end{figure}
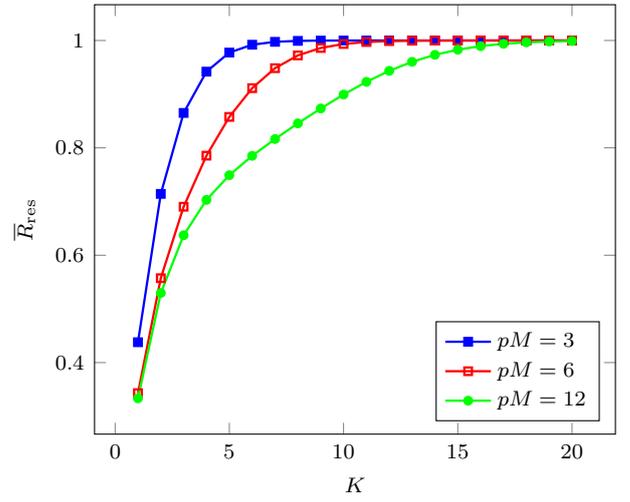

From Theorem~\ref{th:boundsbasic}, we derive results on the expected number of users that is resolved per slot $\barRresolve$, \ie the expected throughput.
\begin{theorem} \label{th:Rresolve}
The expected number of users that is resolved per slot is lower bounded as
\begin{equation}
\barRresolve \geq 1 - \frac{\beta^*-1}{\beta^*(1-q_0)}I_p(K+1,M-K).
\end{equation}
\end{theorem}
\begin{IEEEproof}
We have
\begin{align}
\barRresolve
&= \sum_{L=1}^M \frac{L}{S(L)}\hat q(L) \\
&\geq \sum_{L=1}^K \hat q(L) + \sum_{L=K+1}^M \frac{L}{\beta^*L-1}\hat q(L) \\
&\geq (1-q_0)^{-1}\left( \sum_{L=0}^K q(L) + \frac{1}{\beta^*}\sum_{L=K+1}^M q(L) - q_0 \right) \\
&= 1 - \frac{\beta^*-1}{\beta^*(1-q_0)}I_p(K+1,M-K),
%
\end{align}
where $\hat{q}( L )$ is the probability of having $L \geq 1$ active users and $I_p(K+1,M-K)$ is the regularized incomplete beta function, see~\eqref{eq:BetaFunc}.
\end{IEEEproof}
The result is illustrated in Figure~\ref{fig:avgperf_inK} as a function of $K$ for various values of $p$.


%
%

%
%

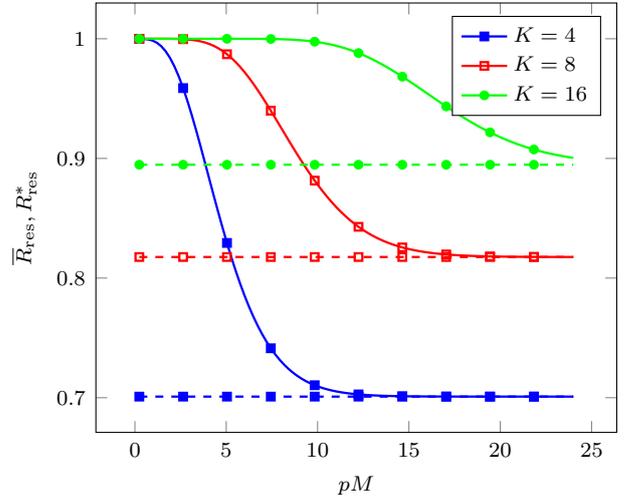
\begin{figure}
\centering
\begin{tikzpicture}
\begin{axis}[
  xlabel=$p M$,ylabel={$\barRresolve, \Rresolve^*$}, 
  font=\scriptsize,
  legend style={
        cells={anchor=west},
        legend pos=north east,
       font=\scriptsize,
    }
]

\addplot[
  line width=.3mm,color=blue, solid,
  mark=square*,mark repeat=10,mark phase=0,mark size=.5mm,mark options={solid}
  ]
table[
  header=false,x index=0,y index=1,
  ]
{matlab_figures/avg_vs_worst.csv};
\addlegendentry{$K=4$};
\addplot[
  line width=.3mm,color=blue, dashed,
 mark=square*,mark repeat=10,mark phase=0,mark size=.5mm,mark options={solid},
  forget plot
  ]
table[
  header=false,x index=0,y index=2,
  ]
{matlab_figures/avg_vs_worst.csv};
\addplot[
  line width=.3mm,color=red, solid,
  mark=square,mark repeat=10,mark phase=0,mark size=.5mm,mark options={solid}
  ]
table[
  header=false,x index=0,y index=3,
  ]
{matlab_figures/avg_vs_worst.csv};
\addlegendentry{$K=8$};
\addplot[
  line width=.3mm,color=red, dashed,
  mark=square,mark repeat=10,mark phase=0,mark size=.5mm,mark options={solid},
 forget plot
  ]
table[
  header=false,x index=0,y index=4,
  ]
{matlab_figures/avg_vs_worst.csv};
\addplot[
  line width=.3mm,color=green, solid,
  mark=*,mark repeat=10,mark phase=0,mark size=.5mm,mark options={solid}
  ]
table[
  header=false,x index=0,y index=5,
  ]
{matlab_figures/avg_vs_worst.csv};
\addlegendentry{$K=16$};
\addplot[
  line width=.3mm,color=green, dashed,
  mark=*,mark repeat=10,mark phase=0,mark size=.5mm,mark options={solid}
  ]
table[
  header=false,x index=0,y index=6,
  ]
{matlab_figures/avg_vs_worst.csv};
\end{axis}
\end{tikzpicture}
\caption{Average and worst case number of users that is resolved per \CS{slot} ($M=1031$). $\Rresolve^*$ in dashed lines, $\barRresolve$ in solid lines. \label{fig:avg_vs_worst}}
\end{figure}

In addition to the expected throughput $\barRresolve$, we are also interested in the worst-case behavior, \ie we analyze $\Rresolve^* = \inf_L \Rresolve(L)$. 
The next result is an immediate corollary of Theorem~\ref{th:boundsbasic}.
 \begin{corollary} \label{cor:Rresolveworst}
 \begin{equation}
 \Rresolve^* \geq \frac{1}{\beta^*}.
 \end{equation}
 \end{corollary}
 We illustrate Corollary~\ref{cor:Rresolveworst} in Figure~\ref{fig:avg_vs_worst}, depicting both $\Rresolve^*$ as well as $\barRresolve$ for various values of $K$.
 The value of $\barRresolve$ is given as a function of $pM$, which is the average number of active users.

\subsection{Net rate in bits per channel use}

Here we consider the net rate $\Rnet(L)$, \ie the overall throughput in bits per channel use that is effectively transmitted.
This performance parameter takes into account the overhead that is generated by the physical-layer network coding, signatures and the tree-splitting. \JGc{In this section we ignore finite block length effects and assume that PLNC can achieve rate $\Rplnc = \frac{1}{2}\log^+_2(P)$ (as given by Theorem~\ref{th:plnc}) with arbitrarily small error probability at a finite block length.}

\JGc{
First, we analyze the number of message bits $D$ that can be transmitted within a block of $N$ channel uses. Let $k$ denote the number of $q$-ary symbols that can be transmitted. We have
\begin{equation}
\frac{k}{N}\log_2 q = \Rplnc. 
\end{equation}
Combining the message bits and the signature we have
\begin{equation}
k=\frac{D}{\log_2q} + N_w,
\end{equation}
where $q=N\log_2 N$ and $N_w$ is given by Theorem~\ref{th:sigcode}. From this we can solve for $D$ as a function of $N$ and evaluate $\Rnet(L) = \frac{D}{N}\Rresolve(L)$. We obtain
\begin{equation}
D = \frac{N}{2}\log_2^+(P) - N_w\log_2(N\log_2 N).
\end{equation}
For completeness, we give the full result in the following corollary.
%
\begin{corollary} \label{cor:Rnet}
The expected number of bits per channel use $\barRnet$ is at least
\begin{multline}
\barRnet \geq \left(\frac{1}{2}\log_2^+(P) - \frac{N_w}{N}\log_2(N\log_2 N)\right) \\
\cdot \left(1 - \frac{\beta^*-1}{\beta^*(1-q_0)}I_p(K+1,M-K)\right),
\end{multline}
where $N_w = \left\lceil\frac{K\log_2 M}{\log_2(N\log N-1) - \log_2 K}\right\rceil + 1$ and $N$ is the block length.
\end{corollary}
}

\JGc{
In the next result we consider the case that $D,N\to\infty$ and consider the maximum of $\barRnet$ over $K$. The result states that the resulting net rate is $\frac{1}{2}\log^+_2(P)$.
\begin{theorem} \label{th:barRnetLimitD}
As $D$ increases, the value of $\barRnet$ optimized over $K$ is $\frac{1}{2}\log^+_2(P)$, \ie
\begin{equation}
\max_{K} \lim_{D\to\infty} \barRnet = \frac{1}{2}\log^+_2(P).
\end{equation}
\end{theorem}
\begin{IEEEproof}
First,
\begin{multline}
\lim_{D\to\infty} \barRnet \\
= \frac{1}{2}\log^+_2(P)\bigg(1 - \frac{\beta^*-1}{\beta^*(1-q_0)}I_p(K+1,M-K)\bigg).
\end{multline}
Now, for $K=M$, the regularized incomplete beta function is at its minimum value $I_p(M+1,0)=0$ and $(\beta^*-1)/\beta^*/(1-q_0)$ is finite. Therefore, the maximum on the right-hand side in the above expression is obtained for $K=M$ and equals $\frac{1}{2}\log^+_2(P)$.
\end{IEEEproof}
}

\JG{
\subsection{Upper Bound}
We consider an upper bound on $\Rnet$ that must be satisfied by any \CS{multiple access} protocol \PP{that serves a batch of arrived packets, each at a different user}:
\begin{theorem} \label{th:upper}
\begin{equation} \label{eq:upper}
\barRnet \leq \sum_{L=1}^M \binom{M}{L}p^L(1-p)^{M-L} \frac{1}{2}\log_2\left(1+LP\right).
\end{equation}
\end{theorem}
\begin{IEEEproof}
The bound is obtained by assuming  \JGb{that all} users and the receiver have complete knowledge about which users are active. Under these assumptions, the problem reduces to a standard Gaussian multi-access channel. The sum rate that can be used by $L$ active users is
\begin{equation}
\Rnet(L) \leq \frac{1}{2}\log_2\left(1+LP\right).
\end{equation}
This immediately leads to \eqref{eq:upper} by taking the expectation over $L$.
\end{IEEEproof}
\JGb{
The following corollary follows directly from Theorem~\ref{th:upper} by an application of Jensen's inequality.
\begin{corollary} \label{cor:upperJensen}
\begin{equation}
\lim_{M\to\infty}\barRnet \leq \frac{1}{2}\log_2(1+pMP).
\end{equation}
\end{corollary}
}
\JGb{In the next subsection we will interpret the upper bound from this section and relate it to the value of $\barRnet$ achieved by the proposed scheme.}
}

%

\subsection{Evaluation}

Figure~\ref{fig:avgperf_inK} shows that as a function of $K$, $\barRresolve$ quickly approaches the maximum value of 1.
This performance parameter is a baseline measure of the efficiency of the random access protocols from the system perspective.
Our results clearly demonstrate the potential of the proposed strategy.
Figure~\ref{fig:avgperf_inK}  also shows that $K$ should increase as the expected number of the active users $pM$ increases, \PP{in order to achieve high throughput}.
Conversely, if $K$ is fixed, the expected throughput drops with $p$ to its lower bound, as shown in Figure~\ref{fig:avg_vs_worst}.

%
%
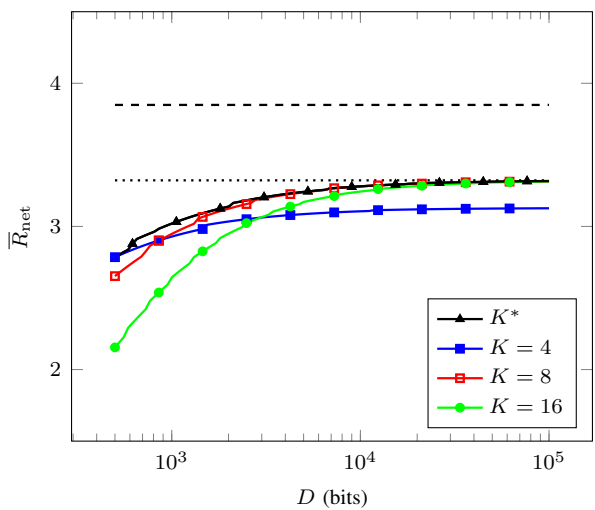
\begin{figure} 
\centering
\begin{tikzpicture}
\begin{semilogxaxis}[
  xlabel=$D$ (bits),ylabel=$\barRnet$, 
   ymin=1.5,ymax=4.5,
  font=\scriptsize,
  legend style={
        cells={anchor=west},
        legend pos=south east,
       font=\scriptsize,
    }
]

\addplot[
  line width=.3mm,color=black, solid,
  mark=triangle*,mark repeat=10,mark phase=5,mark size=.5mm,mark options={solid}
  ]
table[
  header=false,x index=6,y index=7,
  ]
{matlab_figures/Rtotal_inD.csv};
\addlegendentry{$K^*$};
\addplot[
  line width=.3mm,color=blue, solid,
  mark=square*,mark repeat=10,mark phase=0,mark size=.5mm,mark options={solid}
  ]
table[
  header=false,x index=0,y index=1,
  ]
{matlab_figures/Rtotal_inD.csv};
\addlegendentry{$K=4$};
\addplot[
  line width=.3mm,color=red, solid,
  mark=square,mark repeat=10,mark phase=0,mark size=.5mm,mark options={solid}
  ]
table[
  header=false,x index=2,y index=3,
  ]
{matlab_figures/Rtotal_inD.csv};
\addlegendentry{$K=8$};
\addplot[
  line width=.3mm,color=green, solid,
  mark=*,mark repeat=10,mark phase=0,mark size=.5mm,mark options={solid}
  ]
table[
  header=false,x index=4,y index=5,
  ]
{matlab_figures/Rtotal_inD.csv};
\addlegendentry{$K=16$};
\addplot[
  line width=.3mm,color=black, dashed,
  forget plot
  ]
table[
  header=false,x index=0,y index=8,
  ]
{matlab_figures/Rtotal_inD.csv};
\addplot[
  line width=.3mm,color=black, dotted,
  forget plot
  ]
table[
  header=false,x index=0,y index=9,
  ]
{matlab_figures/Rtotal_inD.csv};
\addplot[
  line width=.3mm,color=black, solid,
  mark=triangle*,mark repeat=10,mark phase=5,mark size=.5mm,mark options={solid},
  forget plot
  ]
table[
  header=false,x index=6,y index=7,
  ]
{matlab_figures/Rtotal_inD.csv};
\end{semilogxaxis}
\end{tikzpicture}
\caption{\JGb{The lower bound on $\barRnet$ from Corollary~\ref{cor:Rnet}.
In dotted line the value $1/2\log_2^+(P)$ as given by Theorem~\ref{th:barRnetLimitD}.} ($M=1031$, $pM=3$, $P=10^2$) \label{fig:Rtotal_inD}}
\end{figure}

%
%
\begin{figure}
\centering
\begin{tikzpicture}
\begin{semilogxaxis}[
  xlabel=$D$ (bits),ylabel=$K^*$, 
  font=\scriptsize,
  legend style={
        cells={anchor=west},
        legend pos=north west,
       font=\scriptsize,
    }
]

\addplot[
  line width=.3mm,color=blue, solid,
  ]
table[
  header=false,x index=0,y index=4,
  ]
{matlab_figures/optimalK_inD.csv};
\addlegendentry{$pM=12$};
\addplot[
  line width=.3mm,color=green, solid,
  mark=*,mark repeat=200,mark phase=40,mark size=.5mm,mark options={solid}
  ]
table[
  header=false,x index=0,y index=3,
  ]
{matlab_figures/optimalK_inD.csv};
\addlegendentry{$pM=6$};
\addplot[
  line width=.3mm,color=red, dotted,
  ]
table[
  header=false,x index=0,y index=2,
  ]
{matlab_figures/optimalK_inD.csv};
\addlegendentry{$pM=3$};
\addplot[
  line width=.3mm,color=black, solid,
  mark=square*,mark repeat=200,mark phase=25,mark size=.5mm,mark options={solid}
  ]
table[
  header=false,x index=0,y index=1,
  ]
{matlab_figures/optimalK_inD.csv};
\addlegendentry{$pM=1$};
\end{semilogxaxis}
\end{tikzpicture}
\caption{Value of $K$ that optimizes $\barRnet$, denoted by $K^*$. ($M=1031$, $P=10^2$) \label{fig:OptimalK_inD}}
\end{figure}
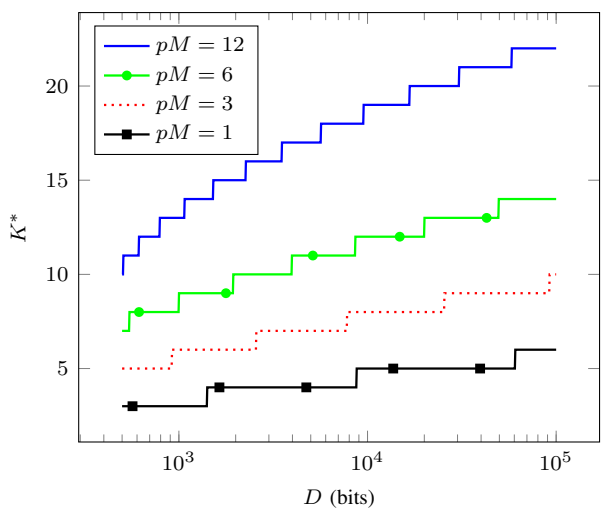

\JGb{In Figure~\ref{fig:Rtotal_inD} we have illustrated our lower bound on $\barRnet$ from Corollary~\ref{cor:Rnet} as a function of $D$, the size \JGc{of} a message transmitted by a user. \JGc{Note that the lowest value for $D$ in the figure is $D=500$, for which at $K=1$ and $P=100$, $N\approx 156$ and we satisfy the required constraint $M\leq q-1 = N\log_2 N -1$ for $M=1031$. The non-smooth behavior of $\barRnet$ is a result of the ceiling operation in the length of the signatures.} In addition to $\barRnet$, Figure~\ref{fig:Rtotal_inD} depicts in a dashed line the upper bound from Theorem~\ref{th:upper}, demonstrating what is the price to pay in information bits per channel use due to: (i) \JGc{using physical-layer network coding to decode a linear combination of messages in each slot,}
(ii) the overhead related to the use of signatures, and (iii) the loss caused by the occurrence of collision and empty slots, compared to the ideal scenario of \JGb{be}forehand knowing the set of active users and using the optimal multi-user code. Figure~\ref{fig:Rtotal_inD} also depicts in a dotted line the value $1/2\log_2^+(P)$. From Theorem~\ref{th:barRnetLimitD} it follows that $1/2\log_2^+(P)$ is the limiting value for $\barRnet$ of our scheme for $D\to\infty$ and $M\to\infty.$ It is clear that no scheme in which one linear combination of messages is decoded at the receiver per slot will be able to achieve a net rate larger than $1/2\log_2^+(P)$. Therefore, Figure~\ref{fig:Rtotal_inD} illustrates that the performance degradation due to (ii) and (iii) diminishes as $D$ increases, such that for already modest values of $D$ in practice, the degradation is mainly due to the \JGc{way in which physical-layer network coding is applied in our approach.}
}

\JGb{This is also illustrated by considering the upper bound from Corollary~\ref{cor:upperJensen}, which gives $\frac{1}{2}\log_2(1+pMP)$, where $pM$ is the expected number of active users, whereas our scheme achieves~$1/2\log_2^+(P)$. A similar behavior was observed in~\cite{goseling2015access_it}, in which the PLNC strategy was extended to allow for several linear combinations to be decoded per slot. Even though this provided a slight improvement, the performance of the resulting strategy did not match the upper bound. Also, using Theorem~\ref{th:upper} itself, instead of Corollary~\ref{cor:upperJensen}, will not close this gap. \JGc{It is an open problem to determine whether the gap to optimality observed in this work and in~\cite{goseling2015access_it} can be reduced by modifying our approach.}
}

\CS{Figure~\ref{fig:Rtotal_inD} also shows that there is an optimal value of $K$ that minimizes the combined overhead of (ii) and (iii), \ie maximizes $\barRnet$ with respect to $D$; the maximum value of $\barRnet$ for the optimal $K$, denoted by $K^*$, is also depicted in Figure~\ref{fig:Rtotal_inD}.}
Finally, Figure~\ref{fig:OptimalK_inD} shows $K^*$ as function of payload length $D$ \JGb{and the expected number of active users.}
We note that finding analytical expressions for $K^*$ involves dealing with partial derivatives of the regularized incomplete beta function and is out of the scope of the paper.





%
%

%
%
\section{Incorporating Successive Interference Cancellation} \label{sec:sic}


In this section we consider an extension of the proposed scheme that includes successive interference cancellation.
Specifically, in Section~\ref{sec:idea}, any \CS{computed sum of the received signals in the slot} that involves more than $K$ users is considered a collision and discarded.
The rationale is that in this case the receiver is not able to determine the set of active users ${\cal L}$ that contribute to the \CS{sum}. 
However, as indicated in Section~\ref{ssec:plnc}, due to the PLNC properties the receiver can successfully receive a linear combination of the packets of all active users.
We show in this section how the receiver can efficiently make use of these collision slots by storing these sums of packets for later use.
In the further text, we refer to this scheme as the SIC-enabled scheme.

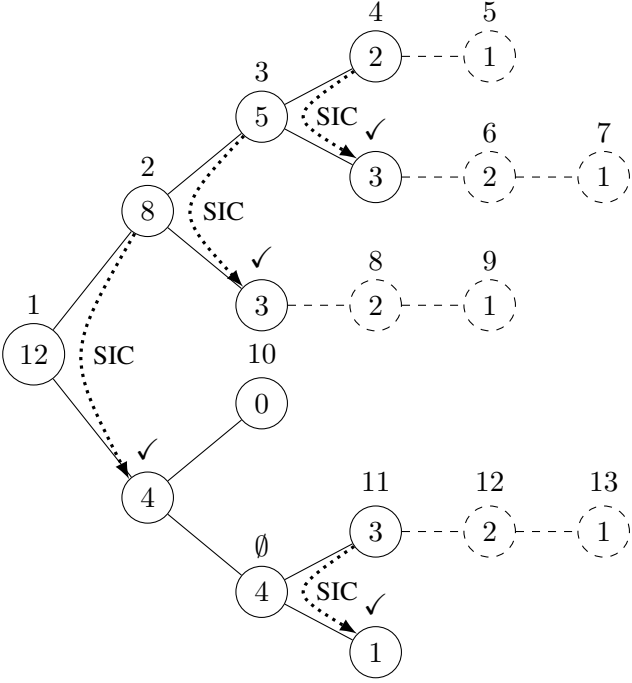
\begin{figure}
\centering
\begin{tikzpicture}[scale=1]
\path[
  every node/.style={circle,draw, label distance=-1mm},
  level distance=15mm, 
  level 1/.style={sibling distance=38mm},
  level 2/.style={sibling distance=25mm},
  level 3/.style={sibling distance=16mm},
]
  node[label=$1$] (r) {$12$}[grow'=right]
    child {node[label=$2$] (a) {$8$}
      child {node[label=$3$] (c) {$5$}
        child {node[label=$4$] (e) {$2$}
          child [dashed] {node[label=$5$] {$1$}
          }
        }
        child {node[label=$\checkmark$] (f) {$3$}
          child [dashed] {node[label=$6$] {$2$}
            child {node[label=$7$] {$1$}
            }
          }
        }
      }
      child {node[label=$\checkmark$] (d) {$3$}
        child [dashed] {node[label=$8$] {$2$}
          child {node[label=$9$] {$1$}
          }
        }
      }
    }
    child {node[label=$\checkmark$] (b) {$4$}
      child {node[label=$10$] {$0$}
      }
      child {node[label=$\emptyset$] (i) {$4$}
        child {node[label=$11$] (g) {$3$}
          child [dashed] {node[label=$12$] {$2$}
            child {node[label=$13$] {$1$}
            }
          }
        }
        child {node[label=$\checkmark$] (h) {$1$}
        }
      }
    };
      
\draw[-latex, very thick, dotted] (a) to[controls=(r.east) and (r.east)] node[right,draw=none] {\small SIC} (b);
\draw[-latex, very thick, dotted] (c) to[controls=(a.east) and (a.east)] node[right,draw=none] {\small SIC} (d);
\draw[-latex, very thick, dotted] (e) to[controls=(c.east) and (c.east)] node[right,draw=none] {\small SIC} (f);
\draw[-latex, very thick, dotted] (g) to[controls=(i.east) and (i.east)] node[right,draw=none] {\small SIC} (h);
\end{tikzpicture}
\caption{Tree splitting with SIC, $K=3$. A checkmark indicates that this signal is obtained by SIC and that no slot is used. The transmission labeled $\emptyset$ is skipped, since it would not bring new information and users are instructed to split.}
\label{fig:treesic}
\end{figure}

\subsection{Description of the SIC-enabled scheme}
The encoder and decoder that are used in each slot are the same as in the proposed scheme of Section~\ref{sec:idea}, \ie encoding is performed as in Section~\ref{ssec:ideaencoder} and decoding as in Section~\ref{ssec:ideadecoder}. The difference is in the action taken by the receiver in case $L>K$, \ie the difference is in the splitting procedure that takes place when more than $K$ users are transmitting.
We proceed by describing the details. 

Recall from Section~\ref{ssec:ideadecoder} that the receiver obtains $\sum_{\ell\in\LL} W_\ell$, which consists of the concatenation of the sums of the signatures $\sum_{\ell\in\LL} W^s_\ell$ and the sums of the codewords $\sum_{\ell\in\LL} W_\ell^d$. If $L > K$ then $\sum_{\ell\in\LL} W^s_\ell$ cannot be uniquely decoded to learn $\LL$.
The key difference between the scheme of Section~\ref{ssec:ideasplit} and the SIC-enabled scheme is that the receiver stores $\sum_{\ell\in\LL} W_\ell$ for future use, instead of discarding it.
\CS{Again,} all users in $\LL$ participate in a splitting protocol with uniform splits into two groups. 
Each user independently of the other users draws a uniformly distributed random number from $\{1,2\}$.
Let $\LL_i$ denote the users with value $i$, $i=1,2$. First, all users with value $1$ enter a new contention resolution phase.
At the end of this phase, which might include recursive splitting steps, the receiver knows $\LL_1$ and $\sum_{\ell\in\LL_1} W_\ell$.
Using the stored information $\sum_{\ell\in\LL} W_\ell$, the receiver now computes
\begin{equation} \label{eq:sicstep}
\sum_{\ell\in\LL_2} W_\ell = \sum_{\ell\in\LL} W_\ell - \sum_{\ell\in\LL_1} W_\ell.
\end{equation}
Note that $\sum_{\ell\in\LL_2} W_\ell$ is exactly the signal that would occur in the first slot of the second subtree in the splitting procedure of Section~\ref{ssec:ideasplit}.
Thus, instead of being obtained through an additional slot, $\sum_{\ell\in\LL_2} W_\ell$ it is obtained through~\eqref{eq:sicstep}.
The corresponding slot is omitted and the contention resolution of users in $\LL_2$ proceeds in the similar fashion.
An overview of the SIC approach is provided in Figure~\ref{fig:treesic}.

An important detail of the scheme that we have not yet discussed is also illustrated in Figure~\ref{fig:treesic}.
Specifically, if $\LL_1=\emptyset$, then $\LL_2=\LL$ and \eqref{eq:sicstep} does not provide new information.
In this case, the first slot in the second subtree can also be omitted and the users are instructed to immediately split again.\footnote{A similar modification to the original, $K=1$ algorithm when $\LL_1=\emptyset$, was proposed in \cite{Massey}. \CS{We also note that the immediate split of the second group, for any value of $|\LL_1|$ and given that $|\LL_2| > K $}, naturally fits into the SIC framework and is the rule, rather than a modification.}
Finally, in case $\LL_1=\LL$, the second phase is (obviously) omitted completely. 

An important difference between our SIC scheme and other SIC-based contention resolution mechanisms \cite{SICTAb,PSLP2014}, is that our approach is based on reliable PLNC, whereas other approaches work on noisy signals. \JG{This has the advantage that it leverages the receiver from the burden to store large quantities of physical-layer output \PP{with a} high precision.}


\subsection{Analysis}

We start with the equivalent of Lemma~\ref{lem:recursion} for the SIC-enabled scheme.
The notation that is used in this section is same the same as Section~\ref{sec:analysis}, with an additional subscript SIC whenever there is a difference with the scheme of Section~\ref{sec:idea}. 
\begin{lemma}
\begin{align}
S_{\textnormal{SIC}} (L) = 
\begin{cases}
1,\quad &\text{if } L=0,\\
L,\quad &\text{if } 1\leq L\leq K,\\
\displaystyle\frac{ 2 \sum_{i=0}^{L-1} p_L ( i ) S_{\textnormal{SIC}} ( i ) }{1 - 2p_L ( L )},\quad &\text{if } L>K.
\end{cases}
\end{align}
\end{lemma}
\begin{IEEEproof}
We focus on the recursive expression, when $L \geq K$ and when a split is performed.
We distinguish three cases, depending on the number of users in the first group after the split.
If there are $1\leq i\leq L-1$ users in the first group, then we require $S_{\text{SIC}}(i)$ slots to resolve this group.
For the second group, the first linear combination of the $L-i$ users is obtained through SIC and $S_{\text{SIC}}(L-i)-1$ additional slots are required.
If there are no users in the first group, then $S_{\text{SIC}}(0)=1$ slot is used for the first group.
For the second group of $L$ users, the first transmission can be omitted, since it is known in advance that it will not provide a new linear combination.
Therefore, $S_{\text{SIC}}(L)-1$ additional slots are required.
Finally, if there are $L$ users in the first group, $S_{\text{SIC}}(L)$ slots will be used for this group and no slots will be used for the second group.
Combining all cases leads to 
\begin{multline} 
\label{eq2}
S_{\text{SIC}} ( L )  = 1 + \sum_{i = 0}^{ L - 1} p_L ( i ) \Big( S_{\text{SIC}} ( i )
+ S_{\text{SIC}} ( L - i ) - 1\Big) \\
+ p_L ( L ) S_{\text{SIC}} ( L ),
\end{multline}
when $L > K$. The proof of the lemma readily follows from \eqref{eq2}, using the facts that $S_{\text{SIC}}( 0 ) = 1$ and $p_L ( i ) = p_L ( L - i ) $, $ i = 0, \dots, L $.
\end{IEEEproof}


The following lemma is the equivalent of Lemma~\ref{lem:TLbound}. The proof is completely analogous to the proof of Lemma~\ref{lem:TLbound} and, therefore, omitted.
\begin{lemma} \label{lem:Ssicbound}
If $\alpha_{\textnormal{SIC}}$ and $\beta_{\textnormal{SIC}}$ satisfy
\begin{align}
\alpha_{\textnormal{SIC}} \leq \gamma_{\textnormal{SIC}} (L) \leq \beta_{\textnormal{SIC}},
\end{align}
for all $ L > K $, where
\begin{align}
\gamma_{\textnormal{SIC}} ( L ) = \frac{ \sum_{i=0}^{K} S_{\text{SIC}} ( i ) p_L ( i ) } { \sum_{ i = 0 }^{ K } p_L ( i ) i  },
\end{align}
then
\begin{align}
\alpha_{\textnormal{SIC}} L \leq S_{\textnormal{SIC}} (L) \leq \beta_{\textnormal{SIC}} L,
\end{align}
for all $L>K$.
\end{lemma}

Next, we present bounds on $\gamma_{\text{SIC}}(L)$.
\begin{lemma}
\begin{align}
\label{eq:S_SIC}
1 \leq \gamma_{\textnormal{SIC}} ( L ) \leq 1 + \frac{1}{(K+1)(2^K - 1 )} .
\end{align}
\end{lemma}
\begin{IEEEproof} Rewrite $\gamma_{\text{SIC}}$ as
\begin{align}
\gamma_{\text{SIC}} ( L ) = \frac{ p_L ( 0 ) + \sum_{i=1}^{K} p_L ( i ) i } { \sum_{ i = 1 }^{ K } p_L ( i ) i  } = 1 + \frac{1}{\sum_{ i = 1 }^{ K } { L \choose i }  i }.
\end{align}
The lower bound trivially holds. Regarding the upper bound, it should be observed that $ { L \choose i } $, $ i = 0, \dots, K $, increases with $L$. Therefore
\begin{align}
\gamma_{\text{SIC}} ( L ) 
\leq&\ \gamma_{\text{SIC}} ( K + 1 ) \\
=&\ 1 +  \frac{1}{\sum_{ i = 1 }^{ K } { K + 1 \choose i }  i } \\
=&\ 1 + \frac{1}{(K+1)(2^K - 1 )}.
\end{align}
\end{IEEEproof}

Based on the above lemma we define
\begin{align}
\alpha_{\text{SIC}}^* &= 1, \\
\beta_{\text{SIC}}^* &= 1 + \frac{1}{(K+1)(2^K - 1 )},
\end{align}
to serve as bounds on $S_{\text{SIC}}(L)$ in Lemma~\ref{lem:Ssicbound}. In Table~\ref{table:TLbounds2} we list some values for $\beta_{\text{SIC}}^*$. Obviously, the upper bound quickly approaches the lower bound of 1 as $K$ increases, implying that $S_{\text{SIC}}(L)$ also quickly approaches $L$, which is the minimal number of slots required to resolve $L$ user transmissions.

%
%
\begin{figure}
\centering
\begin{tikzpicture}
\begin{semilogxaxis}[
  xlabel=$D$ (bits),ylabel=$\barRnet$, 
  font=\scriptsize,
  legend style={
        cells={anchor=west},
        legend pos=south east,
       font=\scriptsize,
    }
]

\addplot[
  line width=.3mm,color=red, solid,
  mark=square*,mark repeat=10,mark phase=8,mark size=.5mm,mark options={solid}
  ]
table[
  header=false,x index=6,y index=7,
  ]
{matlab_figures/Rtotal_inD_SIC.csv};
\addlegendentry{$\overline{R}_{\textnormal{net, SIC}}$};
\addplot[
  line width=.3mm,color=black, solid,
  mark=triangle*,mark repeat=10,mark phase=5,mark size=.5mm,mark options={solid}
  ]
table[
  header=false,x index=6,y index=7,
  ]
{matlab_figures/Rtotal_inD.csv};
\addlegendentry{$\barRnet$};
%
%
%
%
%
%
%
%
%
%
%
%
\addplot[
  line width=.3mm,color=black, dotted,
  forget plot
  ]
table[
  header=false,x index=0,y index=9,
  ]
{matlab_figures/Rtotal_inD_SIC.csv};
\end{semilogxaxis}
\end{tikzpicture}
\caption{Lower bounds on $\barRnet$ and $\overline{R}_{\textnormal{net, SIC}}$ at the optimal value for $K$. \JGb{In dashed line the upper bound on $\barRnet$ from Theorem~\ref{th:upper}. In dotted line the value $1/2\log_2^+(P)$ as given by Theorem~\ref{th:barRnetLimitD}.} ($M=1031$, $pM=3$, $P=10^2$) \label{fig:Rtotal_inD_SIC}}
\end{figure}
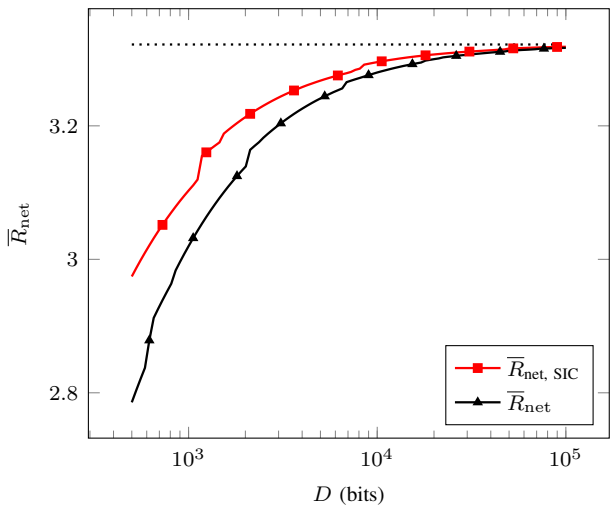

\begin{table}
\centering

\begin{equation*}
\begin{array}{rl}
K &  \beta_{\text{SIC}}^* \\
\hline
1 & 1.5 \\
2 & 1.111\\
4  & 1.036 \\
8 &  1.013 
\end{array}
\end{equation*}

\caption{Values for $\beta^*$ that serve as the upper bound on $S_{\text{SIC}}(L)$.} \label{table:TLbounds2}
\end{table}



We derive results on $\barRresolveSIC$, the expected number of users that is resolved per slot in the SIC-enabled scheme.
The next result is the equivalent of the non-SIC result in Theorem~\ref{th:Rresolve} with $\beta^*$ replaced by $\beta_{\text{SIC}}^*$; the proof is omitted, as it requires only a minor modification to the proof of Theorem~\ref{th:Rresolve}.
\begin{theorem} \label{th:RresolveSIC} The expected number of users that is resolved per slot is lower bounded as
\begin{equation}
\label{eq:R_SIC}
\barRresolveSIC \geq 1 - \frac{\beta_{\textnormal{SIC}}^*-1}{\beta_{\textnormal{SIC}}^*(1-q_0)}I_p(K+1,M-K).
\end{equation}
\end{theorem}
%

The result of Theorem~\ref{th:RresolveSIC} immediately leads to a result on $\overline{R}_{\textnormal{net, SIC}}$, \ie the net rate in bits per channel use that is achievable with SIC.
In Figure~\ref{fig:Rtotal_inD_SIC} we have compared maximum values for $\overline{R}_{\textnormal{net, SIC}}$ and $\barRnet$, both at their individual optimal values of $K$. \JGb{In addition, similar to Figure~\ref{fig:Rtotal_inD}, we have depicted the upper bound from Theorem~\ref{th:upper} and the value $1/2\log_2^+(P)$ as given by Theorem~\ref{th:barRnetLimitD}.}
We observe from Figure~\ref{fig:Rtotal_inD_SIC} that, even though SIC has a significant impact on the value of expected number of slots in a contention period \eqref{eq:S_SIC} and on the expected number of users resolved per slot \eqref{eq:R_SIC}, the impact on the net rate is limited.
\CS{In other words, reusing collision slots in the SIC-enabled scheme only modestly improves net-rate performance.}
We conclude this section by noting that it is straightforward to obtain the other results for the SIC-enabled scheme that correspond to the results in Section~\ref{sec:analysis}.
Therefore, these are not presented here.

%
%
%
%
%

%
%
%
\section{Discussion} \label{sec:discussion}

\JG{In the paper we \PP{have} assumed a unit channel gain model in which users are synchronized to the start of a contention resolution period.
Here we discuss how the assumption of the unit channel gains can be relaxed to take into account channels with fading\PP{, as well as} how the synchronization can be achieved.
The idea is that the start of a contention resolution period is marked by a beacon sent by the base station, \CSm{synchronizing the users}. Upon receiving the beacon, each user that has a message to send estimates its channel to the base station. This estimate is \CSm{also} obtained from the beacon, assuming channel reciprocity. If the channel is sufficiently strong (to be made precise in the following text), then the user becomes active for this contention period, \ie it joins the set of contending users. The active user \emph{inverts} the channel and, during the contention process, \emph{precodes} its transmission by sending the signal $X_m/h_m$, where $h_m$ is the channel coefficient between the $m-$th user and the receiver. This channel, as perceived by the receiver, has unit channel gains. Note that in this case the uncertainty about the users that constitute the set $\LL$ comes both from the sporadic message arrival per user as well as the changes in the channel. We assume a quasi-static fading model, such that $h_m$ stays constant during the contention period, which also implies that the set of active users $\LL$ remains invariant until the contention is resolved. Finally, due to the power constraint, if a user observes a channel with $|h_m|^2<1/P$, it does not become active and does not join the contention set $\LL$. \CS{Such a user} will wait for a next contention period in which it has a stronger channel.
}

\JG{
It is interesting to note that in standard scenarios with PLNC, the channel coefficients are assumed to be known at the receiver~\cite{popovski2007physical, popovski2006anti, nazer11compforw}, which is consistent with the fact that the receiver knows a priori the set of transmitting users; in other words, 
the only issue is to properly select the codebooks and the decoders. However, in our scenario there is uncertainty about the set of transmitting users and it is thus not reasonable to assume that the receiver would know the channel coefficients $\{h_m\}$. Precoding results in a multiple access channel in which the coefficient of each user, as perceived by the receiver, is equal to one. Thus, although the receiver does not know the set of transmitting users, it knows a priori the channel coefficient by which each user is received. With these assumptions, the receiver is able to set up the correct decision regions in order to decode the superposition of the lattice-based codewords from the transmitters.
}

The random-access setup considered in the paper can be categorized as a batch contention resolution, \CSrev{where all users arrive before the start of the contention resolution procedure}, and the contention resolution is performed by the binary fair-splitting of collisions with multiplicity larger than $K$.
The initial collision multiplicity is binomially distributed, as the user arrivals are modeled by Bernoulli trials.
\PP{The sporadic nature of message arrivals is reflected in the fact that $p$ is small, such that \JG{with high probability a} user that is already in $\LL$ does not get a new message during the contention period.} \JG{In order to take into account additional arrivals at users already in $\LL$, we could extend the protocol by allowing a user to indicate with an additional bit in their message that it has one more message. Such a user would be granted another collision-free transmission once it succeeds to send its first message successfully.} 

\JGb{In the paper t}he focus of the analysis \JGb{of the proposed scheme} was on the derivation of the bounds on the protocol performance, characterizing the behavior as $K$ \PP{grows, \ie as a larger number of users becomes resolvable simultaneously.} The obtained bounds are given in closed form, depend only on $K$ and provide a direct insight into the scheme's performance.
We note that, in principle, one could reuse the results from \cite{MF1985} and \cite{SICTAb}, and obtain non-recursive expressions for the expected duration of the contention resolution period given that the initial collision multiplicity is $L$, for the original $S(L)$ and SIC-enabled variant $S_{\text{SIC}}(L)$, respectively.
Specifically, the difference to the analysis carried out in \cite{MF1985} and \cite{SICTAb} is in the generalization in the range of the initial conditions to $L = 0,1,\dots,K$.
However, taking into account that these expressions for $S(L)$/$S_{\text{SIC}}(L)$ will inevitably be in a convolved form\footnote{Cf. (3.31) in \cite{MF1985} and (30) in \cite{SICTAb}.}, and that the bounds that have been derived in the current paper become increasingly tight as $K$ grows, we have not pursued obtaining non-recursive formulae for $S(L)$/$S_{\text{SIC}}(L)$.


The straightforward generalization of the proposed scheme is to assume $Q$-ary splitting and investigate the optimal values of $Q$ and the optimal splitting probabilities, including the potential exploitation of the fact that the collision multiplicity $L$ is always known.
However, our preliminary analysis shows that the gain that could be achieved by such optimization is negligible compared to the gain that is achieved by optimizing $K$.

Further extensions could include more elaborate arrival models, \CSrev{where new users can become active after the start of the (current) contention resolution period, which can be handled in the blocked (gated), windowed and free-access manner \cite{Massey,SICTAb}}.
In this regard, we conjecture that, based on the results from \cite{Massey}, the original variant of the proposed scheme will experience improvements w.r.t. the worst case performance, see Figure~\ref{fig:avg_vs_worst}, if the windowed access is used.
On the other hand, we conjecture that the blocked access for the SIC-enabled variant is optimal.
Specifically, in this variant of the scheme, all non-empty slots that occur during the contention resolution period are useful, which eliminates the arguments for `bounding' the initial collision multiplicity that fosters the windowed access.
We also note that the analogous result is formally derived in \cite{SICTAb} for the case $K = 1$.

We further remark that Figures~\ref{fig:Rtotal_inD} and \ref{fig:OptimalK_inD} clearly \JGb{illustrate the performance} loss due to the access protocol elements, \ie physical layer network coding, signature coding and the contention resolution mechanism, w.r.t. the ideal scenario in which set of the active users and their identities are a-priori known and the optimal multi-user code is used.
Such a comparison is omitted \JGb{in related work, which consistently relies on} idealized assumptions regarding the physical layer and \JGb{does not take the overhead related to user identification explicitly into account.}

\CS{As shown \JGb{in this paper,} the dominant loss as the length of the data portion of user packet grows is due to PLNC; it is currently an open problem if this loss is an inherent property of PLNC or an artifact of the computation coding construction that is developed in~\cite{nazer11compforw}.
}

\JGf{The current work has not addressed the complexity of decoding the signatures. It would be of interest to design low-complexity decoding algorithms by basing the construction of the signature codes on, for instance, BCH codes (instead of Lindstr\"om's construction) and by using existing decoding techniques as done in~\cite{bar1993forward, ordentlich2017low}.}
%

As already mentioned in the introduction, the mechanisms introduced \JGb{in} this paper through signature\JGb{s} and physical-layer network coding are \JGb{not} limited to tree-splitting algorithms. \CSrev{In particular, PLNC and signature coding were also applied in the context of coded random access \cite{SVGP2016}, where the advantage is that the feedback comes less often compared to the tree-splitting protocols.
However, the proposed protocol offers several advantages over coded random access.
One advantage is related to the sensitivity of the performance to the choice of parameters, which in both cases depend on the number of active users.
In this respect, the proposed protocol can provide meaningful performance without a priori knowledge of the number of active users and the optimal selection of $K$, as demonstrated in Figure~\ref{fig:avg_vs_worst}.
On the other hand, the performance of coded random access is much more sensitive, with a chance of performing poorly if the parameters are not selected optimally, c.f. Figure~5 in \cite{L2011} that shows that the throughput becomes very low if the frame has not been dimensioned properly.
Further, the use of feedback, which drives the resolution process in the proposed scheme, enables the receiver to start resolving users after a number of slots that scales logarithmically with the initial collision multiplicity.
On the other hand, in frame-based coded random access, the resolution process is initiated only at the end of the frame; similarly, frameless coded random access exhibits the well known threshold effect in terms of user resolution, when most of the users become resolved at the end of the contention period \cite{PSLP2014}.
In this respect, it can be argued that the proposed protocol provides a more balanced profile of the user resolution delay}. \JGc{It will be of interest to compare the performance of our proposed strategy with work on models in which the user detection problem is related to compressed sensing, as done in, for instance,~\cite{chen2014many}. An important difference between~\cite{chen2014many} and the current work is that in~\cite{chen2014many} the number of users scales with the block length.}

\PP{One relevant extension of this work is directed to the practical scenario in which physical-layer network coding is not carried out using lattices, but rather a practical symbol-by-symbol modulation combined with error-control coding. This fact, along with the finite block length, necessarily leads to a nonzero probability of errors at the receiver, which requires suitable modification of the protocol. \JGc{Another problem of practical relevance is to develop efficient algorithms that decode sums of signatures. Finally, an} interesting extension is related to the }\CSrev{construction of signatures when the assumption of the equal powers at the point of reception does not hold.}

\bibliographystyle{IEEEtran}
\bibliography{IEEEabrv,signatures}

\end{document}